\documentclass[aps,prc,amsfonts,preprintnumbers,superscriptaddress,showpacs,nofootinbib]{revtex4}

\usepackage{graphics}
\usepackage{psfrag}
\usepackage{epsfig}
\usepackage{epsf}
\usepackage{float}
\usepackage{amsmath,amsthm,amssymb,scrtime,fancyhdr,totpages,bbm,empheq,units,mathtools,placeins,xcolor}
\usepackage{bbm}
\allowdisplaybreaks[1]

\newcommand{\beq}{\begin{equation}}
\newcommand{\eeq}{\end{equation}}
\newcommand{\beqa}{\begin{eqnarray}}
\newcommand{\eeqa}{\end{eqnarray}}
\newcommand{\nn}{\nonumber \\ }

\newcommand{\Lag}{\mathcal{L}}

\newcommand{\tp}{\vec{\tau}\cdot \vec{\pi}}

\newcommand{\one}{\mathbbm{1}}

\begin{document}

\title{Two-nucleon electromagnetic current in chiral effective field
  theory: \\ one-pion exchange and short-range contributions}

\author{S.~K\"olling}
\email[]{Email: s.koelling@fz-juelich.de}
\affiliation{Institut f\"ur Theoretische Physik II, Ruhr-Universit\"at Bochum, D-44780 Bochum, Germany}
\affiliation{Helmholtz-Institut f\"ur Strahlen- und Kernphysik (Theorie)
and Bethe Center for Theoretical Physics,\\
 Universit\"at Bonn, D-53115 Bonn, Germany}
\author{E.~Epelbaum}
\email[]{Email: evgeny.epelbaum@rub.de}
\affiliation{Institut f\"ur Theoretische Physik II, Ruhr-Universit\"at Bochum, D-44780 Bochum, Germany}
\author{H.~Krebs}
\email[]{Email: h.krebs@fz-juelich.de}
\affiliation{Institut f\"ur Theoretische Physik II, Ruhr-Universit\"at Bochum, D-44780 Bochum, Germany}
\author{U.-G.~Mei{\ss}ner}
\email[]{Email: meissner@hiskp.uni-bonn.de}
\affiliation{Helmholtz-Institut f\"ur Strahlen- und Kernphysik (Theorie)
and Bethe Center for Theoretical Physics,\\
 Universit\"at Bonn, D-53115 Bonn, Germany}
\affiliation{Institut f\"ur Kernphysik (IKP-3), Institut for Advanced
  Simulation (IAS-4) and J\"ulich Center for Hadron Physics, \\
  Forschungszentrum J\"ulich,   D-52425 J\"ulich, Germany}
\date{\today}

\begin{abstract}
We derive the leading one-loop contribution to the one-pion exchange and
short-range two-nucleon electromagnetic current operator in the framework of chiral effective field
theory. The derivation is carried out using the method of unitary
transformation. Explicit results for the current and charge densities are given in
momentum and coordinate space. 

\end{abstract}


\maketitle


\section{Introduction}
\def\theequation{\arabic{section}.\arabic{equation}}
\setcounter{equation}{0}
\label{sec:intro}

There have been several recent studies on the nuclear exchange electromagnetic
currents within the framework of chiral effective field theory 
\cite{Pastore:2008ui,Pastore:2009is,Kolling:2009iq,Pastore:2011ip}, see
also \cite{Park:1995pn} for an older calculations which, however, is limited to the
near-threshold kinematics. These studies constitute  a natural extension to
photon-induced reactions of the theoretical framework formulated by Weinberg 
two decades ago \cite{Weinberg:1991um}, see \cite{Epelbaum:2008ga} for a recent review
article.  To derive the exchange currents from
the most general effective chiral Lagrangian the authors of
Refs.~\cite{Pastore:2008ui,Pastore:2009is,Pastore:2011ip} used the framework of ``old-fashioned''
time-ordered perturbation theory along the lines of
\cite{Weinberg:1991um}. This approach leads, in general, to explicitly
energy-dependent potentials and currents. Such  energy dependence might cause
difficulties in few-body applications. To obtain energy-independent 
nuclear potentials we employed in Refs.~\cite{Epelbaum:1998ka,Epelbaum:1999dj}
the method of unitary transformation. In Ref.~\cite{Kolling:2009iq}, we applied this
approach to the long-range parts of the leading two-pion exchange
contributions to the current and charge densities. In this manuscript, we
derive all remaining contributions to the two-nucleon current and charge
densities at the leading loop order (i.e.~of order $eQ$ with $Q\sim M_\pi$ referring to
low external momenta). 

It is important to emphasize conceptual differences between our work and the one 
by Pastore et al.~\cite{Pastore:2008ui,Pastore:2009is,Pastore:2011ip}. These authors 
limit themselves to deriving the momentum dependence of the one-pion exchange current and charge 
operators at the leading loop level in chiral effective field theory \emph{without considering 
renormalization}. Consequently, the values of the various  low-energy constants (LECs)
entering their expressions \emph{cannot} be taken from other sources such as e.g.~pion-nucleon 
scattering. One, therefore, looses one of the greatest strengths of the effective field theory approach,
namely 
the ability to relate different processes. The calculation presented in our work is more 
ambitious aiming at the derivation of \emph{renormalized} expressions for the exchange 
current and charge operators. This is a highly nontrivial task for the one-pion exchange 
contributions. Contrary to the calculations in the Goldstone boson 
and single-nucleon sectors, one is dealing here only with an irreducible part of the 
amplitude (giving rise to nuclear forces and currents) which itself is not an observable 
quantity and is affected by unitary transformations. On the other hand, there is no freedom 
in absorbing the divergences generated by the loop corrections to the one-pion exchange operators
since all $\beta$-functions of the corresponding LECs are fixed and well known. 
As we will demonstrate in this work, it is indeed possible to exploit the above mentioned unitary ambiguity 
in such a way that all divergences emerging from pion loops are indeed absorbed by redefinition 
of the LECs leading to the finite result for the current and charge 
operators, where the values of renormalized LECs can be taken from other sources. 

Our manuscript is organized as follows. In section \ref{sec1} we provide a short
summary of the method of unitary transformation (UT) and explain very briefly 
the adopted power counting scheme. The effective Lagrangian employed in our
calculation is specified in section \ref{sec2}. The results for various
contributions to the one-pion exchange current and charge densities are
discussed in detail in section \ref{sec3}. Section \ref{sec10} deals
with the derivation of the short-range contributions. 
A comparison between our work and the calculations by Pastore et al.~is 
presented in section \ref{sec_pastore}.
The results of our work are summarized in section
\ref{sec_summ}. The expressions for the relevant terms in
the effective pion-nucleon-photon Hamiltonian density are listed in 
appendix \ref{append_Ham}, while appendix
\ref{app2} collects the expressions for the relevant loop integrals. 
The Expression for the current and charge density in configuration space are given in appendix~\ref{app3}.

\section{Anatomy of the calculation}
\def\theequation{\arabic{section}.\arabic{equation}}
\setcounter{equation}{0}
\label{sec1}

The derivation of the electromagnetic nuclear current operators is carried out along the lines of
Ref.~\cite{Kolling:2009iq}, 
see also \cite{Epelbaum:2002gb}. The main steps are summarized
below. 
\begin{itemize}
\item
We begin with the effective chiral Lagrangian in the heavy-baryon formulation
and express it in terms of \emph{renormalized} pion and nucleon 
fields and apply the canonical formalism along the lines of Ref.~\cite{Gerstein:1971fm} to
derive the corresponding Hamilton density. 
The contributions from tadpole diagrams are taken into account by performing normal
ordering of the resulting Hamilton density. 
Notice that the terms in the effective
Lagrangian/Hamiltonian involving two and more insertions of an external 
electromagnetic field $\mathcal{A}^\mu$ are not taken into account since we restrict ourselves to the
one-photon-approximation (however, the method can straightforwardly be generalized to 
two-photon processes such as Compton scattering off light nuclei).  
The obtained contributions to the Hamilton density
are listed in appendix \ref{append_Ham}.
\item
To decouple the purely nucleonic subspace of the Fock space from the rest we
apply an appropriately chosen UT  
\beq
\tilde H \equiv U^\dagger H U = \left( \begin{array}{cc} \eta \tilde H \eta  &
    0 
\\ 0 & \lambda \tilde H \lambda \end{array} \right)\,. 
\eeq
Here, $\eta$ ($\lambda$)  denote 
projection operators onto the purely nucleonic (the remaining) part of the
Fock space satisfying $\eta^2 = \eta$, $\lambda^2 = \lambda$, $ \eta \lambda 
= \lambda \eta = 0$ and $\lambda + \eta = {\bf 1}$. The resulting nuclear
Hamiltonian $\eta \tilde H \eta$ gives rise to the chiral potentials in
Refs.~\cite{Epelbaum:1999dj,Epelbaum:2004fk,Epelbaum:2006eu,Bernard:2007sp}. 
Both the UT and the transformed Hamiltonian 
are calculated by making a perturbative expansion in powers of $Q/\Lambda$, 
with $Q$ and $\Lambda$ referring to the soft and
hard scales of the order of the pion and $\rho$-meson masses, respectively. 
The power counting is most easily formulated in terms of the 
canonical field dimension $\kappa$ of the interaction vertices, 
\begin{equation}
\label{pow_fin}
H = \sum_{\kappa = 1}^{\infty} H^{(\kappa )} \,,
\quad \quad \kappa_i = d_i + \frac{3}{2} n_i + p_i - 4\,.
\end{equation}
Here, $d_i$, $n_i$ and
$p_i$ refer to the number of derivatives or $M_\pi$-insertions, nucleon field
and pion field operators, respectively. The explicit form of the strong part
of the unitary operator $U$, i.e.~the one in the absence of the external
electromagnetic field, sufficient to derive the nuclear force up to N$^3$LO is
given in Ref.~\cite{Epelbaum:2007us}.   
\item
The effective nuclear current operator $\eta J^\mu
( x ) \eta$ acting in the purely nucleonic subspace of the Fock space 
is defined according to \cite{Eden:1995rf,Kolling:2009iq}
\beq
\label{current_def}
J^\mu ( x )  =  \eta  \, U^\dagger
J^\mu_{\rm bare} ( x ) U \, \eta  \,.
\eeq
Here, $J^\mu_{\rm bare} (x)$ denotes the hadronic current density which enters the
effective Lagrangian $\mathcal{L}_{\pi N
  \gamma}$ describing the interaction of pions and nucleons with an
external electromagnetic field $\mathcal{A}^\mu$. It is given by 
\beq
J^\mu_{\rm bare} (x) = \partial_\nu \frac{\partial \mathcal{L}_{\pi N \gamma}}{\partial
  (\partial_\nu \mathcal{A}_\mu)} - \frac{\partial \mathcal{L}_{\pi N \gamma}}{\partial
  \mathcal{A}_\mu}\,.
\eeq
The $\lambda$-components of the effective current operator do not 
need to be taken into account as long as one stays below the pion production threshold.
The above definition of $\eta J^\mu ( x ) \eta$ does, in fact, not fully
incorporate the freedom in the choice of UT. In
particular, one can introduce $\eta$-space UTs 
$\eta U' \eta$ that depend explicitly on the external electromagnetic field 
$\mathcal{A}_\mu$ such that 
\beq
\eta U' \eta \, \Big|_{\mathcal{A}_\mu = 0} = 1_\eta . 
\eeq
Applying such UTs on the nuclear Hamiltonian $\eta \tilde H \eta$ will
generate further contributions to the nuclear current operator. 
The resulting ambiguity is analogous to the one in the strong sector which  
is described in detail in Ref.~\cite{Epelbaum:2006eu,Epelbaum:2007us}. 
As will be shown below, renormalizability 
of the one-pion exchange contributions at the one-loop level strongly
restricts the ambiguity in the definition of $\eta J^\mu ( x ) \eta$. 
\item
The final step in the derivation involves evaluating the emerging loop integrals and
expressing the current operator in terms of renormalized low-energy constants. This is carried
out within the framework of dimensional regularization (DR) which allows us to adopt
the known expressions for the $\beta$-functions of the LECs entering
$\mathcal{L}_{\pi N}^{(3)}$.  
\end{itemize}
In the following sections, the various steps in the derivation
of the current will be discussed in detail.

\section{Effective Lagrangian}
\def\theequation{\arabic{section}.\arabic{equation}}
\setcounter{equation}{0}
\label{sec2}

In this work we employ the standard heavy-baryon formulation for the effective
Lagrangian. The terms needed in the calculation of the leading loop
corrections to the one-pion exchange and short-range current operator 
read
\cite{Gasser:1983yg,Bellucci:1994eb,Ecker:1995rk,Fettes:1998ud,Fettes:2000gb,Epelbaum:2000kv,Gasser:2002am}
\begin{eqnarray}
\label{lagr}
\Lag_{\pi\pi}^{(2)} & = & \frac{F^2}{4}\langle D_\mu U D^\mu U^\dagger +
\chi_+ \rangle \, , \nn
\Lag_{\pi\pi}^{(4)} & = & \frac{l_3}{16}\langle \chi_+ \rangle^2 +
\frac{l_4}{16}\biggl( 2\langle D_\mu U D^\mu U^\dagger\rangle \langle\chi_+
\rangle + 2 \langle \chi^\dagger U \chi^\dagger U + \chi U^\dagger \chi
U^\dagger \rangle -4 \langle \chi^\dagger \chi\rangle\biggr) \nn
&+& i \frac{l_6}{2}\langle f_{\mu\nu}^{\rm R} D^\mu U D^\nu U^\dagger
+
f_{\mu\nu}^{\rm L} \left(D^\mu U\right)^\dagger D^\nu U\rangle  + \ldots \,,\nn
\Lag_{\pi N}^{(1)} & = & \bar{N}_v \left[i \left(v \cdot D \right) +
  \mathring{g}_A \left(S\cdot u \right) \right]N_v \,,\nn
\Lag_{\pi N}^{(2)} & = & \bar{N}_v \biggl[\frac{1}{2\mathring{m}}\left(v\cdot D
  \right)^2-\frac{1}{2\mathring{m}}\left(D\cdot D
  \right)
 - i \frac{\mathring{g}_A}{2\mathring{m}} \{S\cdot D, v\cdot u\} + \ldots
 \biggr]N_v \,,\nn
\Lag_{\pi N}^{(3)} & = & \bar{N}_v \biggl[d_{16} S\cdot u \langle\chi_+ \rangle
  + i d_{18} S^\mu \left[D_\mu,\chi_- \right]
  +\tilde{d}_{28} \left(i \langle\chi_+ \rangle v\cdot D +
    \textrm{h.c.}\right)+ d_6 v^\nu \left[ D^\mu,\tilde{f}_{\mu\nu}^+\right]
      + d_7 v^\nu \left[D^\mu, \langle f_{\mu\nu}^+\rangle \right] \nn
&+& d_7 v^\nu \left[D^\mu, \langle f_{\mu\nu}^+\rangle \right]
      + d_8 \epsilon^{\mu\nu\alpha\beta}v_\beta\langle \tilde{f}_{\mu\nu}^+
      u_\alpha \rangle 
      + d_9 \epsilon^{\mu\nu\alpha\beta}v_\beta\langle f_{\mu\nu}^+
       \rangle u_\alpha
      + d_{20} iS^\mu v^\nu \left[ \tilde{f}_{\mu\nu}^+, v\cdot u\right]
      + d_{21} i S^\mu \left[\tilde{f}_{\mu\nu}^+,u^\nu \right] \nn
&+& d_{22} S^\mu \left[D^\nu, f_{\mu\nu}^- \right]\biggr] N_v  + \ldots\,,\nn
\mathcal{L}_{NN}^{(0)} & = & -\frac{1}{2}C_S \bar{N}_v N_v \, \bar{N}_v N_v
  +2 C_T \bar{N}_v S_\mu N_v \, \bar{N}_v S^\mu N_v \,, \nn
\mathcal{L}_{NN}^{(2)} & = &\frac{1}{2}\alpha_1 \left[(\bar{N}_v
      \overrightarrow{D}_\mu N_v )( \bar{N}_v
      \overrightarrow{D}^\mu N_v ) + \textrm{h.c.} \right]
+ \alpha_2 (\bar{N}_v \overrightarrow{D}_\mu N_v )( \bar{N}_v
   \overleftarrow{D}^\mu N_v ) + \alpha_3 (\bar{N}_v N_v )(\bar{N}_v (
    \overleftarrow{D}^2 +  \overrightarrow{D}^2 )N_v ) \nn
&+& \alpha_4 
(\bar{N}_v N_v )( \bar{N}_v
  \overleftarrow{D}_\mu\overrightarrow{D}^\mu N_v )
+ \frac{i}{2}\alpha_5
\epsilon_{\mu\nu\rho\sigma}v^\mu\left[(\bar{N}_v
    \overrightarrow{D}^\nu N_v )(\bar{N}_v
    \overleftarrow{D}^\rho  S^\sigma N_v ) -\textrm{h.c.} \right]
+ i \alpha_6 \epsilon_{\mu\nu\rho\sigma}v^\mu (\bar{N}_v N_v ) \nn
&\times&
(\bar{N}_v\overleftarrow{D}^\nu  S^\rho
  \overrightarrow{D}^\sigma  N_v )
 + i\alpha_7 \epsilon_{\mu\nu\rho\sigma} v^\mu (\bar{N}_v S^\nu N_v
)(\bar{N}_v \overleftarrow{D}^\rho
  \overrightarrow{D}^\sigma N_v )
+ \frac{i}{2}\alpha_8 \epsilon_{\mu\nu\rho\sigma} v^\mu\left[
  (\bar{N}_v \overrightarrow{D}^\nu N_v
  )(\bar{N}_v S^\rho \overrightarrow{D}^\sigma N_v ) -  \textrm{h.c.}   \right]\nn
&+& \frac{1}{2}\left(\alpha_9 g_{\mu\rho}g_{\nu\sigma} + \alpha_{10}
  g_{\mu\sigma}g_{\nu\rho} + \alpha_{11}
  g_{\mu\nu}g_{\rho\sigma}\right)
 \left[(\bar{N}_vS^\rho \overrightarrow{D}^\mu
    N_v )(\bar{N}_vS^\sigma \overrightarrow{D}^\nu 
  N_v ) + \textrm{h.c.}  \right]\nn
&+& \left( \alpha_{12} g_{\mu\rho}g_{\nu\sigma} + \alpha_{13}
  g_{\mu\sigma}g_{\nu\rho} + \alpha_{14} g_{\mu\nu}g_{\rho\sigma}
\right)  (\bar{N}_v S^\rho \overrightarrow{D}^\mu N_v
)(\bar{N}_v \overleftarrow{D}^\nu S^\sigma N_v )\nn
&+& \frac{1}{2}\left(\frac{1}{2}\alpha_{15}\left(g_{\mu\rho}g_{\nu\sigma} +
    g_{\mu\sigma}g_{\nu\rho} \right) +
  \alpha_{16}g_{\mu\nu}g_{\rho\sigma}\right)
 \left[(\bar{N}_v
    \overleftarrow{D}^\mu S^\rho
    \overrightarrow{D}^\nu N_v )(\bar{N}_v
   S^\sigma N_v ) +  \textrm{h.c.}    \right]\nn
&+& \frac{1}{2}\left(\frac{1}{2}\alpha_{17}\left(g_{\mu\rho}g_{\nu\sigma} +
    g_{\mu\sigma}g_{\nu\rho} \right) +
  \alpha_{18}g_{\mu\nu}g_{\rho\sigma}\right)
(\bar{N}_v(\overleftarrow{D}^\mu \overleftarrow{D}^\nu  +
    \overrightarrow{D}^\mu \overrightarrow{D}^\nu ) S^\rho  N_v
)(\bar{N}_vS^\sigma  N_v ) \nn
&+&  \, \epsilon_{\mu\nu\rho\sigma}v^\mu f^{\nu\rho} \left[
   L_1 \left(\bar{N}_vS^\sigma \tau^3N_v\bar{N}_vN_v - \bar{N}_vS^\sigma
      N_v\bar{N}_v\tau^3N_v \right) + L_2
    \bar{N}_vS^\sigma N_v\bar{N}_vN_v \right] + \ldots \, ,
\end{eqnarray}
where $v$ denotes the nucleon four-velocity, $\langle \ \rangle$ stands for the trace in
the flavor space and the spin vector is defined as
\beq
S_\mu  =  \frac{i}{2}
\gamma_5 \sigma_{\mu\nu}v^\nu \,, \quad \sigma_{\mu\nu} =
\frac{i}{2}[\gamma_\mu,\gamma_\nu]\,, \quad
\{S_\mu,S_\nu\}  =  \frac{1}{2}\left(v_\mu v_\nu - g_{\mu\nu} \right) \,, \quad
[S_\mu,S_\nu] = i \epsilon_{\mu\nu\rho\sigma}v^\rho S^\sigma \,,
\eeq
with the last two relations holding in four dimensions. Further, $F$,
$\mathring{m}$ and $\mathring{g}_A$ refer to the pion decay constant, nucleon
mass and the nucleon axial-vector coupling in the chiral limit while $l_i$,
$d_i$, $C_{S,T}$, $\alpha_i$ and $L_{1,2}$  are further LECs. Notice that 
we only list those terms in the effective Lagrangian which are explicitly
needed in our calculations. For example, we omit all terms in $\Lag_{\pi N}^{(2)}$
proportional to the LECs $c_i$ as they lead to vertices with at least two
pions\footnote{The only exception is the $c_1$-term which also has a
  contribution that does not involve pion field operators. This contribution
  can be absorbed into redefinition of the nucleon mass.} and thus will not
contribute to the current operator up to the leading-loop order.  
We further emphasize that the terms in  $\mathcal{L}_{NN}^{(2)}$ do not
correspond to the minimal set, see \cite{Epelbaum:2000kv,Girlanda:2010ya} for
more details and relations 
between the different $\alpha_i$. We will address this issue and list the
minimal set of contact interactions the  nucleon rest-frame at the end of this
section. 
The superscript $i$ in
$\Lag_{\pi N}^{(i)}$,  $\Lag_{\pi \pi}^{(i)}$ and $\Lag_{NN}^{(i)}$ refers to the number of
derivatives and/or quark mass insertions. The unitary $2 \times 2$ matrix $U$
parametrizes the Goldstone Boson fields and is given by 
\beq
U  =  \one +i \frac{\tp}{F} - \frac{\pi^2}{2F^2} - i \xi \frac{\pi^2
    \tp}{F^3} + \frac{(8\xi-1)}{8F^4}\pi^4 + \mathcal{O} (\pi^6) \,,
\eeq
where $\xi$ is a constant representing the freedom in the definition of the
pion fields.  The popular $\sigma$-model gauge and exponential parametrization of the
matrix $U$
correspond to $\xi = 0$ and $\xi =1/6$, respectively. Notice that
physical observables calculated  using the effective
Lagrangian are, clearly, independent on a particular parametrization of $U$.
The quantity $\chi_+$ is defined via
\beq
\chi_+ = u^\dagger \chi u^\dagger + u \chi^\dagger u\,, \quad \quad
\chi = 2 B \mathcal{M} \equiv M^2 \one_2
\eeq
with $B$ and  $\mathcal{M} =
\mbox{diag} (m_u ,\, m_d )$  being a constant and the light quark matrix
accounts for the explicit chiral symmetry breaking and gives rise to the pion
mass  
\beq
M_\pi^2 = M^2 \left( 1 + \mathcal{O} (\mathcal{M}^2 ) \right)\,.
\eeq
The covariant derivatives of the pion and nucleon fields are defined by 
\beqa
D_\mu U &=& \partial_\mu U - i r_\mu U + i U \ell_\mu \,, \nn
u_\mu & = & i \left[ u^\dagger \left(\partial_\mu-i r_\mu \right) u - u \left(
    \partial_\mu -i \ell_\mu\right)u^\dagger\right] \,, \nn
D_\mu N_v & = & \left[\partial_\mu +\Gamma_\mu -i v_\mu^{\rm (s)} \right] N_v ~,
\nn
\Gamma_\mu  &=&  \frac{1}{2} \left[u^\dagger (\partial_\mu - i r_\mu) u + u
  (\partial_\mu -i \ell_\mu) u^\dagger \right]\,,
\eeqa
where $r_\mu$, $l_\mu$ and $v_\mu^{\rm (s)}$ denote the  external right-,
left-handed and isoscalar vector currents, respectively, and $u =
\sqrt{U}$. The derivative operators $\overrightarrow{D}_\mu$ and
$\overleftarrow{D}_\mu$ entering $\Lag_{NN}^{(i)}$ are defined via
\beq
\bar{N}_v \overrightarrow{D}_\mu N_v = \bar{N}_v \left(
   \partial_\mu N_v\right) + \bar{N}_v \left( \Gamma_\mu - i
   v_\mu^{(s)}\right) N_v\,,\quad \quad 
\bar{N}_v \overleftarrow{D}_\mu N_v = \left( \partial_\mu \bar{N}_v \right)
   N_v - \bar{N}_v \left( \Gamma_\mu - i
   v_\mu^{(s)}\right) N_v\,.
\eeq
Further, $f_{\mu\nu}^{\rm L,R}$ and  $v_{\mu\nu}^{(s)}$
denote the field strength tensors associated with external left-, 
right-handed and the isoscalar currents, 
\beq
f_{\mu\nu}^{\rm R} = \partial_\mu r_\nu - \partial_\nu r_\mu - i \left[ r_\mu,
  r_\nu \right]\,, \quad \quad 
f_{\mu\nu}^{\rm L} = \partial_\mu l_\nu - \partial_\nu l_\mu - i \left[ l_\mu,
  l_\nu \right]\,,\quad \quad 
v_{\mu\nu}^{(s)} = \partial_\mu v_\nu^{(s)} - \partial_\nu v_\mu^{(s)}\,,
\eeq
while the corresponding covariantly transforming quantities $f_{\mu\nu}^{\pm}$
which enter the pion-nucleon Lagrangian are defined according to
\beq
f_{\mu\nu}^{\pm} = u^\dagger \left( f_{\mu\nu}^{\rm R} +  v_{\mu\nu}^{(s)}
\right) u \pm  u\left( f_{\mu\nu}^{\rm L} +  v_{\mu\nu}^{(s)}
\right) u^\dagger\,.
\eeq
We also used traceless matrices $\tilde f_{\mu\nu}^\pm$ defined according to  $\tilde f_{\mu\nu}^\pm\equiv
f_{\mu\nu}^\pm  - \langle  f_{\mu\nu}^\pm  \rangle/2$. In this work, we are
interested in describing the coupling to an external electromagnetic field. In
that case, the left- and
right-handed  currents $r_\mu$ and $l_\mu$ and the isoscalar current
$v_\mu^{(s)}$ have to be chosen as
\beq
r_\mu = \ell_\mu = e  \frac{\tau^3}{2} \mathcal{A}_\mu, \quad v_\mu^{\rm
  (s)} = e  \frac{\mathcal{A}_\mu}{2},
\eeq
where $\mathcal{A}_\mu$ refers to the electromagnetic four-potential.

We now turn our attention to the Lagrangians $\Lag_{NN}^{(0,2)}$ involving four nucleon field
operators. At the order considered, there is no
need to account for terms involving pion fields.  
Notice further that Poincar\'e covariance implies that only
7 out of 18 constants $\alpha_i$  are independent and, in addition, also
determines the coefficients in front of the leading $1/m_N^2$ corrections to
contact terms, see \cite{Epelbaum:2000kv,Pastore:2009is,Girlanda:2010ya} for
more details and explicit expressions. In the power counting scheme we adopt
in the present work, the nucleon mass is treated as a heavier scale compared
to the breakdown scale of the chiral expansion, see Refs.~\cite{Weinberg:1991um,Kolling:2009iq} for more
details. Accordingly, there is no need to take into account the leading 
relativistic $1/m_N^2$-corrections to the short-range two-nucleon current at the
order we are working.  Switching to the rest-frame of the nucleon with
$v_\mu=(1,0,0,0)$,  making use
of the partial integrations and incorporating constraints due to the Galilean
invariance allows to express the Lagrangian for contact interactions 
in the standard basis in terms of $C_{1,
  \ldots ,7}$ used e.g. in \cite{Ordonez:1995rz,Epelbaum:1998ka,Epelbaum:1999dj}:
\begin{eqnarray}
  \mathcal{L}_{NN}^{(2)} & = & - \frac{1}{2}C_1 \Big[ (
      N^\dagger \vec{\nabla}N
    )^2 +  N^\dagger \vec{\nabla} N \cdot 
  \vec{\nabla}N^\dagger  N   + \textrm{h.c.}\Big] 
 + \frac{1}{4}C_2 \Big[N^\dagger N  N^\dagger
    \vec{\nabla}^2N  + N^\dagger \vec{\nabla} N \cdot 
  \vec{\nabla}N^\dagger  N   + \textrm{h.c.}\Big]\nn
&+& \left(\frac{1}{2}C_3 \delta_{ij}\delta_{kl}+ \frac{1}{4}C_6 \left(  \delta_{ik}\delta_{jl} +
      \delta_{il}\delta_{kj}\right)\right) \left(  \nabla_i N^\dagger \sigma_k \nabla_j N
    +   \nabla_i \nabla_j N^\dagger
     \sigma_k  N  + \textrm{h.c.}   \right) \left( N^\dagger \sigma_l  N \right)\nn
&+& \left(\frac{1}{8}C_4\delta_{ij}\delta_{kl} +\frac{1}{16} C_7 \left(  \delta_{ik}\delta_{jl} +
      \delta_{il}\delta_{kj}\right) \right) \biggl[ N^\dagger \sigma_k  \nabla_i N \nabla_j N^\dagger
  \sigma_l  N  + \nabla_i \nabla_j N^\dagger
     \sigma_k  N N^\dagger \sigma_l  N   + \textrm{h.c.} 
   \biggr]\nn
&+&¸\frac{i}{8}C_5 \biggl[  N^\dagger \vec{\nabla} N \cdot
  \vec{\nabla}N^\dagger \times\vec{\sigma} N  + 
    \vec{\nabla}N^\dagger N \cdot
   N^\dagger \vec{\sigma}\times \vec{\nabla}N 
- N^\dagger N  \vec{\nabla}N^\dagger \cdot
  \vec{\sigma}\times \vec{\nabla}N  
+ N^\dagger \vec{\sigma}N \cdot 
  \vec{\nabla}N^\dagger \times \vec{\nabla}N  \biggr] \nn
&-& \frac{i}{4}C_2 \, e \, \vec{\mathcal{A}}\cdot \biggl[ N^\dagger \hat{e} N 
      N^\dagger \overleftrightarrow{\nabla} N   - N^\dagger N 
 N^\dagger \hat{e} \overleftrightarrow{\nabla}  N   \biggr]
- \frac{i}{4}e\, \left(C_4 \delta_{ij}\delta_{kl}+ \frac{1}{2}C_7 \left(  \delta_{ik}\delta_{jl} +
      \delta_{il}\delta_{kj}\right)\right) \nn
&& {} \times \biggl[ N^\dagger \sigma_k \nabla_i N
  N^\dagger\sigma_l \hat{e} N \mathcal{A}_j -2
N^\dagger \sigma_k \hat{e} N \nabla_j N^\dagger \sigma_l  N \mathcal{A}_i 
 + \nabla_i N^\dagger\hat{e}\sigma_k  N N^\dagger \sigma_l  N
\mathcal{A}_j \biggr]\nn
&-& \frac{1}{8}C_5 \, e  \, \vec{\mathcal{A}}\cdot \!\! \biggl[
\left( N^\dagger \vec{\nabla}N  {+} \vec{\nabla}N^\dagger N
  \right) \times
 N^\dagger \hat{e}  \vec{\sigma} N 
 {+} N^\dagger  \hat{e} N    
  \left(\vec{\nabla}N^\dagger \times \vec{\sigma}N  {-}
  N^\dagger \vec{\sigma} \times \vec{\nabla}  N \right)
+ N^\dagger N \nn
&& {} \times  \left(N^\dagger \hat{e} \vec{\sigma}\times\vec{\nabla} N 
  {-} \vec{\nabla}N^\dagger \times \vec{\sigma} \hat{e} N \right) {+}  
N^\dagger \vec{\sigma}N
\times  \left( \vec{\nabla}N^\dagger\hat{e} N  {-} N^\dagger   \hat{e}  \vec{\nabla}N \right)
\biggr] 
\nn
&-& e \, \vec{\nabla} \times \vec{A} \cdot \left[  L_1 \left(N^\dagger
    \vec{\sigma} \, \tau^3 N N^\dagger N -
    N^\dagger \vec{\sigma} N N^\dagger \,\tau^3 N \right)  + L_2 N^\dagger \vec{\sigma}
  N N^\dagger N 
  \right] + \ldots
\end{eqnarray}
where we have introduced 
\beq
\hat{e}  =  \frac{\one  + \tau^3}{2}\, , \quad \quad 
N^\dagger\, \overleftrightarrow{\nabla} \,N= N^\dagger \vec{\nabla} \, N -
N^\dagger\, \overleftarrow{\nabla}\, N\,.
\eeq 
Notice that we only kept terms at most linear in the electromagnetic
four-potential. 

As already pointed out in the previous section, the derivation of the exchange
current operator is carried out using the method of unitary transformation
which requires the knowledge of the Hamilton density and the Noether
currents. The transition from the
Lagrangian to the Hamiltonian is achieved employing the standard canonical
formalism. An extended discussion on this can be found in 
Refs.~\cite{Gerstein:1971fm,Epelbaum:2002gb,Kolling:2009iq}. All terms 
in the resulting Hamilton density which enter the calculation are listed in
appendix \ref{append_Ham}.

\section{One-pion exchange current}
\def\theequation{\arabic{section}.\arabic{equation}}
\setcounter{equation}{0}
\label{sec3}

We now turn to the derivation of the two-nucleon electromagnetic current due to
a single pion exchange. In section \ref{sec4}, the derivation and explicit
results for the leading loop contributions are presented. Tree-level
contributions and the renormalization are considered in sections \ref{sec5} and
\ref{sec6}, respectively. Next, in section \ref{sec7} we discuss the leading
relativistic corrections. Final results for the one-pion exchange current and
charge density both in momentum and coordinate spaces
are summarized in  section \ref{sec8}. 

\subsection{Loop contributions}
\def\theequation{\arabic{section}.\arabic{equation}}
\label{sec4}

\begin{figure}[tb]
\vskip 1 true cm
  \includegraphics[width=0.9\textwidth,keepaspectratio,angle=0,clip]{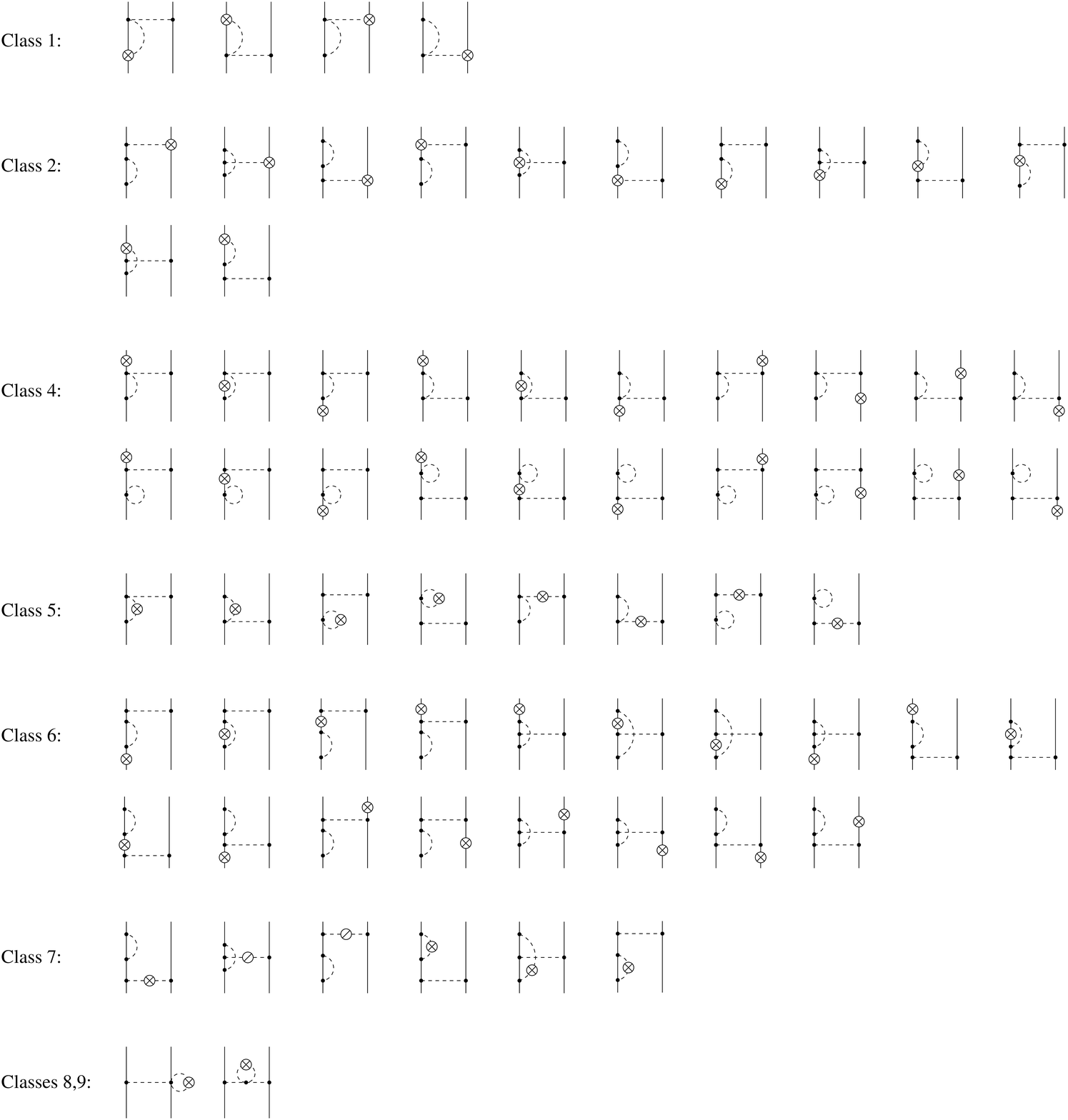}
  \caption{Leading loop contributions to the one-pion exchange current
    operator. Solid and dashed lines refer to nucleons and pions,
    respectively. Solid dots are the lowest-order vertices from the effective
    Lagrangian while the circle-crosses represent insertions of the electromagnetic
    vertices as explained in the text. Diagrams resulting from interchanging
    the nucleon lines are not shown.}
  \label{fig:loops}
\end{figure}

Following Ref.~\cite{Kolling:2009iq}, we classify  various loop contributions according to the powers of
the LEC $g_A$ and the type of the hadronic current $J^\mu_{20}$, $J^\mu_{21}$
or $J^\mu_{02}$ as shown in  Fig.~\ref{fig:loops}. Here and in what follows, we adopt the notation of
Refs.~\cite{Epelbaum:2007us,Kolling:2009iq}. In particular, the subscripts $a$ and $b$ in
$H_{ab}^{(\kappa)}$ and ${J^\mu_{ab}}^{(\kappa)}$ refer to the
number of the nucleon and pion fields, respectively, while the
superscript $\kappa$ gives the dimension of the operator defined in
Eq.~(\ref{pow_fin}). 
Notice that while class-$3$ terms proportional to $g_A^0$ and involving an
insertion of $J^\mu_{02}$ contribute to the two-pion exchange current, they 
do not generate one-pion exchange diagrams. On the other hand we now have additional 
contributions from class-$8,9$ terms which do not contribute to two-pion exchange diagrams 
and, for that reason, were not considered in Ref.~\cite{Kolling:2009iq}. 
We further emphasize that, strictly, speaking (i.e.~according to the power of $g_A$), 
these diagrams belong to class 5. 
The algebraic structure of the current operator in Fock space in terms of 
$H_{ab}$ and $J^\mu_{ab}$ for seven classes is given in appendix A 
of Ref.~\cite{Kolling:2009iq}. The new terms corresponding to the classes 8
and 9  have the form:
\begin{eqnarray}
  J_{\rm c8} & = & \eta\left[
           H_{21}^{(1)} \frac{\lambda^1}{E_\pi} H_{23}^{(3)} \frac{\lambda^2}{E_\pi} J_{02}^{(-1)}
         +H_{21}^{(1)} \frac{\lambda^1}{E_\pi} J_{02}^{(-1)} \frac{\lambda^3}{E_\pi} H_{23}^{(3)}
         +H_{23}^{(3)} \frac{\lambda^3}{E_\pi} H_{21}^{(1)} \frac{\lambda^2}{E_\pi} J_{02}^{(-1)}
          \right]\eta +
  \textrm{h.c.}\,, \nn 
  J_{\rm c9} & = & -\eta\biggl[ H_{21}^{(1)} \frac{\lambda^1}{E_\pi}
  H_{21}^{(1)} \frac{\lambda^2}{E_\pi} H_{04}^{(2)} \frac{\lambda^2}{E_\pi}
  J_{02}^{(-1)} 
          + H_{21}^{(1)} \frac{\lambda^1}{E_\pi} H_{04}^{(2)}
          \frac{\lambda^3}{E_\pi} H_{21}^{(1)} \frac{\lambda^2}{E_\pi}
          J_{02}^{(-1)} 
          + H_{04}^{(2)} \frac{\lambda^4}{E_\pi} H_{21}^{(1)}
          \frac{\lambda^3}{E_\pi} H_{21}^{(1)} \frac{\lambda^2}{E_\pi}
          J_{02}^{(-1)}\\
          &&{}+ H_{21}^{(1)} \frac{\lambda^1}{E_\pi} H_{21}^{(1)}
          \frac{\lambda^2}{E_\pi} J_{02}^{(-1)} \frac{\lambda^4}{E_\pi}
          H_{04}^{(2)} 
          + H_{21}^{(1)} \frac{\lambda^1}{E_\pi} H_{04}^{(2)}
          \frac{\lambda^3}{E_\pi} J_{02}^{(-1)} \frac{\lambda^1}{E_\pi}
          H_{21}^{(1)} 
          + H_{21}^{(1)} \frac{\lambda^1}{E_\pi} J_{02}^{(-1)}
          \frac{\lambda^3}{E_\pi} H_{21}^{(1)} \frac{\lambda^4}{E_\pi}
          H_{04}^{(2)}\biggr]\eta+ 
  \textrm{h.c.} \,. \nonumber
\end{eqnarray}
Here, the superscript $i$ of $\lambda^i$ refers to the number of pions in the
corresponding intermediate state. Further, $E_\pi$ denotes the total energy of
pions in the corresponding state, $E_\pi = \sum_i \sqrt{\vec l_i\, ^2 +
  M_\pi^2}$, with $\vec l_i$ the corresponding pion momenta.   
We remind the reader that the representation for the power counting in terms
of the canonical dimension $\kappa$ allows one  to easily read off the chiral
order associated to a given contribution by simply adding together the dimensions
$\kappa$  of $H_{ab}^{(\kappa)}$ and ${J^\mu_{ab}}^{(\kappa)}$. 

As already pointed out in section \ref{sec1} and in
Ref.~\cite{Kolling:2009iq}, we have  
to employ additional UTs in the $\eta$-space in order to 
maintain renormalizability of the one-pion exchange contributions, see
Refs.~\cite{Epelbaum:2007us} for a related discussion. For the case at hand,
one can distinguish between the strong UTs and the ones depending on the
electromagnetic four-potential $\mathcal{A}$. The general form of the strong
UTs up to the considered order in the chiral expansion is given in
Ref.~\cite{Epelbaum:2007us}.  These continuous UTs are parametrized in terms
of some (a-priori arbitrary) ``angles'' $\bar\alpha_i$.\footnote{In that reference, the angles were 
denoted by $\alpha_i$.}
These parameters turn out to be strongly
constrained if one requires that matrix elements of the resulting nuclear
potentials can be made finite by means of redefinition of certain LECs,
i.e.~if one demands renormalizability at the level of the nuclear
Hamiltonian. For the UTs considered in Ref.~\cite{Epelbaum:2007us}, this
condition was shown to lead to a unique expression for the four-nucleon force
which does not depend on $\bar\alpha_i$ any more. Similarly, the expressions for
the two-pion exchange current operator obtained in Ref.~\cite{Kolling:2009iq} are also
$\bar\alpha_i$-independent. The additional electromagnetic UTs have not been
discussed in that reference as they turned out not to affect the two-pion
exchange contributions. As will be shown below, it is necessary to employ 
such additional UTs to maintain renormalizability of the one-pion
exchange current. To be specific, we consider the $\eta$-space UT of the form  
\beq
\label{UTbeta}
U = e^S
\eeq
where $S$ is an anti-hermitian operator acting in the $\eta$-space, $S=\eta S
\eta$, $S^\dagger = - S$. At the order considered, this operator can be
parametrized as 
\beq
\label{betas}
S = \sum_{i=1}^7  \bar\beta_i S_i 
\eeq
with $\bar\beta_i$ being arbitrary constants and 
\beqa
S_1 &=& \eta \left[ J_{02}^{(-1)} \frac{\lambda^2}{E_\pi^2} H_{22}^{(2)} \; - \;
  H_{22}^{(2)} \frac{\lambda^2}{E_\pi^2} J_{02}^{(-1)} \right] \eta\,, \nn[3pt]
S_2 &=& \eta \left[ H_{21}^{(1)} \frac{\lambda^1}{E_\pi^2} J_{20}^{(-1)}
  \frac{\lambda^1}{E_\pi} H_{21}^{(1)}\; - \;
  H_{21}^{(1)} \frac{\lambda^1}{E_\pi} J_{20}^{(-1)} 
\frac{\lambda^1}{E_\pi^2} H_{21}^{(1)} \right]\eta\,, \nn[3pt]
S_3 &=& \eta \left[ J_{20}^{(-1)} \eta H_{21}^{(1)}
  \frac{\lambda^1}{E_\pi^3} H_{21}^{(1)}\; - \;
  H_{21}^{(1)} \frac{\lambda^1}{E_\pi^3} H_{21}^{(1)}\eta 
 J_{02}^{(-1)} \right]\eta\,, \nn[3pt]
S_4 &=& \eta\left[ J_{02}^{(-1)} \frac{\lambda^2}{E_\pi^2}H_{21}^{(1)}
    \frac{\lambda^1}{E_\pi}H_{21}^{(1)} - H_{21}^{(1)}
    \frac{\lambda^1}{E_\pi}H_{21}^{(1)} \frac{\lambda^2}{E_\pi^2}J_{02}^{(-1)}
  \right]\eta  \,, \nn[3pt]
S_5 &=& \eta\left[ J_{02}^{(-1)} \frac{\lambda^1}{E_\pi}H_{21}^{(1)}
    \frac{\lambda^1}{E_\pi^2}H_{21}^{(1)} - H_{21}^{(1)}
    \frac{\lambda^1}{E_\pi^2}H_{21}^{(1)} \frac{\lambda^2}{E_\pi}J_{02}^{(-1)}
  \right]\eta  \,, \nn[3pt]
S_6 &=& \eta\left[ H_{21}^{(1)} \frac{\lambda^1}{E_\pi}J_{02}^{(-1)} 
    \frac{\lambda^1}{E_\pi^2}H_{21}^{(1)} - H_{21}^{(1)}
    \frac{\lambda^1}{E_\pi^2}J_{02}^{(-1)} \frac{\lambda^2}{E_\pi}H_{21}^{(1)}
  \right]\eta  \,, \nn[3pt]
S_7 &=& \eta\left[ H_{21}^{(1)} \frac{\lambda^1}{E_\pi^3}J_{21}^{(0)} - J_{21}^{(0)}
    \frac{\lambda^1}{E_\pi^3}H_{21}^{(1)}
  \right]\eta  \,.
\label{unitS}
\eeqa
The action of these UTs onto the one-pion exchange contribution to the
lowest-order effective Hamilton operator,  
\beq
H^{(0)} = \eta \left[ H_{20}^{(2)} + H_{40}^{(2)} - H_{21}^{(1)}
  \frac{\lambda^1}{E_\pi} H_{21}^{(1)} \right]\,,
\eeq
with $H_{20}^{(2)}$ denoting the nonrelativistic kinetic energy term,
induces additional, $\bar\beta$-dependent class-2, class-5, class-6 and class-7 contributions:
\beqa
 \delta J_{\rm c2} & = & \beta_7 \, \eta \left[
            H_{21}^{(1)} \frac{\lambda^1}{E_\pi} H_{21}^{(1)} \eta
            H_{21}^{(1)} \frac{\lambda^1}{E_\pi^3} J_{21}^{(0)}  
          - H_{21}^{(1)} \frac{\lambda^1}{E_\pi} H_{21}^{(1)} \eta
          J_{21}^{(0)} \frac{\lambda^1}{E_\pi^3} H_{21}^{(1)}  
        \right]\eta + \textrm{h.c.}\,, \nn       
\delta J_{\rm c5} & = & \bar\beta_1\, \eta\left[H_{21}^{(1)}
  \frac{\lambda^1}{E_\pi} H_{21}^{(1)}  \eta J_{02}^{(-1)} 
 \frac{\lambda^2}{E_\pi^2}  H_{22}^{(2)}  - 
H_{21}^{(1)}  \frac{\lambda^1}{E_\pi} H_{21}^{(1)} \eta H_{22}^{(2)} \frac{\lambda^2}{E_\pi^2} J_{02}^{(-1)}
  \right]\eta + \textrm{h.c.} \,, \nn
\delta  J_{\rm c6} & = & \bar\beta_2\,   \eta \left[ H_{21}^{(1)} \frac{\lambda^1}{E_\pi}
  H_{21}^{(1)} \eta H_{21}^{(1)} \frac{\lambda^1}{E_\pi^2} J_{20}^{(-1)}
  \frac{\lambda^1}{E_\pi} H_{21}^{(1)} - H_{21}^{(1)}
  \frac{\lambda^1}{E_\pi^2} J_{20}^{(-1)} \frac{\lambda^1}{E_\pi} H_{21}^{(1)}
  \eta H_{21}^{(1)} \frac{\lambda^1}{E_\pi} H_{21}^{(1)} \right]\eta + 
  \textrm{h.c.} \, , \nn
 &+& \bar\beta_3\, \eta\left[ H_{21}^{(1)} \frac{\lambda^1}{E_\pi^3}
  H_{21}^{(1)} \eta J_{20}^{(-1)} \eta H_{21}^{(1)} \frac{\lambda^1}{E_\pi}
  H_{21}^{(1)} -  \eta H_{21}^{(1)} \frac{\lambda^1}{E_\pi} H_{21}^{(1)} \eta
  H_{21}^{(1)} \frac{\lambda^1}{E_\pi^3} H_{21}^{(1)} \eta J_{20}^{(-1)}
\right]\eta + \textrm{h.c.} \, , \nn 
\delta  J_{\rm c7} & = & \bar\beta_4\, \eta\biggl[H_{21}^{(1)}
\frac{\lambda^1}{E_\pi}H_{21}^{(1)} \eta J_{(02)}^{(-1)} \frac{\lambda^2}{E_\pi^2} H_{21
}^{(1)} \frac{\lambda^1}{E_\pi} H_{21}^{(1)} - H_{21}^{(1)}
\frac{\lambda^1}{E_\pi}H_{21}^{(1)} \eta H_{21}^{(1)} \frac{\lambda^1}{E_\pi}
H_{21}^{(1)} \frac{\lambda^2}{E_\pi^2}J_{02}^{(-1)}\biggr]\eta +
  \textrm{h.c.} \, , \nn
&+& \bar\beta_5\, \eta\biggl[H_{21}^{(1)}
\frac{\lambda^1}{E_\pi}H_{21}^{(1)} \eta J_{02}^{(-1)} \frac{\lambda^2}{E_\pi} H_{\pi
N} \frac{\lambda^1}{E_\pi^2} H_{21}^{(1)} - H_{21}^{(1)}
\frac{\lambda^1}{E_\pi}H_{21}^{(1)} \eta H_{21}^{(1)} \frac{\lambda^1}{E_\pi^2}
H_{21}^{(1)} \frac{\lambda^2}{E_\pi}J_{02}^{(-1)} \biggr]\eta +
  \textrm{h.c.} \, , \nn
&+& \bar\beta_6\, \eta\biggl[H_{21}^{(1)}
\frac{\lambda^1}{E_\pi}H_{21}^{(1)} \eta H_{21}^{(1)}
\frac{\lambda^1}{E_\pi^2}J_{02}^{(-1)}\frac{\lambda^1}{E_\pi}H_{21}^{(1)} - H_{21}^{(1)}
\frac{\lambda^1}{E_\pi}H_{21}^{(1)} \eta H_{21}^{(1)}
\frac{\lambda^1}{E_\pi}J_{02}^{(-1)}\frac{\lambda^1}{E_\pi^2}H_{21}^{(1)} \biggr]\eta +
  \textrm{h.c.} \, .
\eeqa  
It turns out to be convenient to express $\bar\beta_{4,5,6}$ in terms of another
set of constants $\beta$, $\gamma$ and $\delta$ defined as:
\beq
 \bar \beta_4 \equiv -\beta + \delta\, , \qquad 2\bar\beta_5 \equiv - \beta + \gamma\,,
 \qquad 2\bar\beta_6 \equiv -\beta -\gamma\, .
\eeq
Already at this stage we emphasize that three of the seven parameters, namely 
$\gamma$, $\bar\beta_2$ and $\bar\beta_7$,  do not affect the leading one-loop
contributions to the one-pion exchange
matrix elements. 

After these preliminary remarks, we are now in the position to discuss the
results for the one-loop contributions. Here and in what follows, the
expressions for a class-$X$ contribution $J_{\rm cX}^\mu$ refer to the matrix element
defined according to  
\beq
\label{notation}
\langle \vec p_1 {}'\, \vec p_2 {}' | J^\mu | \vec p_1 \, \vec p_2 \rangle 
= \delta (\vec p_1 {}' + \vec p_2 {}' - \vec p_1 - \vec p_2 - \vec k ) 
\, \left[ J_{\rm cX}^\mu \, + \, (1 \leftrightarrow
 2) \right]\,. 
\eeq
Here and in what follows, $\vec p_i$ ($\vec p_i\, '$) refers to the initial
(final) momentum of the nucleon $i$.   
We will also frequently use the momentum transfer variables 
$q_{1,2} \equiv \vec p_{1,2}\,' -   \vec p_{1,2}$. 
The expressions for the two-pion exchange current and charge densities were
given in Ref.~\cite{Kolling:2009iq} in terms of the most general set of
spin-momentum vector and scalar operators $\vec O_{1 \ldots 24}$ and 
$O_{1 \ldots 8}^S$ as well as isospin operators $T_{1 \ldots 5}$. 
We found that this representation leads to unnecessarily involved expressions
in the case of the one-pion exchange and short-range currents. We, therefore, refrain
from using the operators $\vec O_{1 \ldots 24}$, $O_{1 \ldots 8}^S$ and $T_{1
  \ldots 5}$ in the present work. 
 
Evaluating matrix elements of the operators in the Fock space as discussed
above, we obtain the following results for the matrix elements of the current
density:
\begin{eqnarray}
\label{current_dens}
  \vec{J}_{\rm c1} & = & - e\frac{\mathring{g}_A^2 i}{16F^4}\left[\vec{\tau}_1\times\vec{\tau}_2\right]^3 \, \vec{\sigma}_1
 \frac{\vec{\sigma}_2\cdot\vec{q}_2}{q_2^2 + M_\pi^2} \int
 \frac{d^3l}{(2\pi)^3} \frac{1}{\omega_l}\,, \nn
 \vec{J}_{\rm c2}  & = &
 e\frac{\mathring{g}_A^4 i}{6F^4}\left[\vec{\tau}_1\times\vec{\tau}_2\right]^3 \, \vec{\sigma}_1
 \frac{\vec{\sigma}_2\cdot\vec{q}_2}{q_2^2 + M_\pi^2}  \int
 \frac{d^3l}{(2\pi)^3} \frac{l^2}{\omega_l^3} \,,\nn
\vec{J}_{\rm c5} &  = &
e\frac{\mathring{g}_A^2 \, i}{32F^4}\left[\vec{\tau}_1\times\vec{\tau}_2\right]^3\,
\frac{\vec{\sigma}_2\cdot\vec{q}_2}{q_2^2 + M_\pi^2} \, \int
\frac{d^3l}{(2\pi)^3} \vec{l} \,  \frac{\vec{l}\cdot\vec{\sigma}_1}{\omega_+
  \omega_- (\omega_+ + \omega_-)}  \nn
&& {} -\left( 1 - \bar\beta_1\right)e\frac{\mathring{g}_A^2 \,
  i}{16F^4}\left[\vec{\tau}_1\times\vec{\tau}_2\right]^3
\frac{\vec{\sigma}_2\cdot\vec{q}_2}{q_2^2 + M_\pi^2} \, \vec{\sigma}_1
\cdot\vec{q}_1  \int \frac{d^3l}{(2\pi)^3}\, \vec{l}
\, \frac{\omega_- -  \omega_+}{\omega_+ \omega_- (\omega_+ + \omega_-)^2}\,,
\nn 
   \vec{J}_{\rm c7} & = & - e\frac{\mathring{g}_A^4
i}{4F^4}\,  \left[\vec{\tau}_1\times\vec{\tau}_2\right]^3
\left( \vec{q}_1 - \vec{q}_2\right)\frac{\vec{\sigma}_1\cdot\vec{q}_1}{q_1^2 +
  M_\pi^2}\frac{\vec{\sigma}_2\cdot\vec{q}_2}{q_2^2 + M_\pi^2} \, \frac{1}{3} \int \frac{d^3l}{(2\pi)^3} \frac{l^2}{\omega_l^3}\,\nn
&&{}+  e\frac{\mathring{g}_A^4
i}{8F^4}\left[\vec{\tau}_1\times\vec{\tau}_2\right]^3
\frac{\vec{\sigma}_2\cdot\vec{q}_2}{q_2^2 + M_\pi^2}\, \int
 \frac{d^3l}{(2\pi)^3} \,\vec{l} 
\, \left( 
  \vec{l}\cdot\vec{q}_2
   \vec{\sigma}_1\cdot \vec{k}-  \vec{k}\cdot\vec{q}_2
   \vec{\sigma}_1 \cdot\vec{l} 
 \right)
  \frac{\omega_+^2 + \omega_+\omega_-  + \omega_-^2}{\omega_+^3 \omega_-^3 (\omega_+ +
  \omega_-)} 
\nn
&&{}-e\frac{\mathring{g}_A^4
i}{32F^4}\left[\vec{\tau}_1\times\vec{\tau}_2\right]^3
\frac{\vec{\sigma}_2\cdot\vec{q}_2}{q_2^2 + M_\pi^2}\,
\vec{\sigma}_1\cdot\vec{q}_2 \,\int
\frac{d^3l}{(2\pi)^3} \,\vec{l} \,  
  (k^2-l^2)\,\biggl[ 2( \beta -1)\frac{(\omega_- - \omega_+)(\omega_+^2 + 3\omega_+
      \omega_- + \omega_-^2)}{\omega_+^3 \omega_-^3(\omega_+ + \omega_-)^2}
    \nn
&&{}+
    \delta \frac{(\omega_- - \omega_+)(\omega_-^2 + \omega_+^2)}{\omega_-^3
      \omega_+^3 (\omega_+ + \omega_-)^2}\biggr] \,,
    \nn 
\vec{J}_{\rm c8} & = &  -e\frac{\mathring{g}_A^2 \,
    i}{32F^4}\left[\vec{\tau}_1\times\vec{\tau}_2\right]^3 \frac{\vec{\sigma}_2\cdot\vec{q}_2}{q_2^2 + M_\pi^2} \int
  \frac{d^3l}{(2\pi)^3} 
 \vec{l} \ \frac{ \vec{l}\cdot\vec{\sigma}_1}{\omega_+ \omega_- (\omega_+  +
    \omega_- )}\,, \nn
    \vec{J}_{\rm c9} & = & 
e\frac{\mathring{g}_A ^2 \,
      i}{32F^4}\left[\vec{\tau}_1\times\vec{\tau}_2\right]^3 \frac{\vec{\sigma}_1 \cdot \vec{q}_1}{q_1^2 +
      M_\pi^2}\frac{ \vec{\sigma}_1 \cdot\vec{q}_2}{q_2^2 + M_\pi^2} \int
    \frac{d^3l}{(2\pi)^3} \vec{l} \  \frac{\vec{l}\cdot\left( \vec{q}_1 -
      \vec{q}_2\right)}{\omega_+ \omega_- (\omega_+ + \omega_-)}\, ,
\end{eqnarray}
and the charge density:
\begin{eqnarray}
\label{charge_dens}
\rho_{\rm c6} & = & 
e\frac{\mathring{g}_A^4
}{4F^4}\, \frac{1}{3} \, \tau_2^3 \, 
\vec{\sigma}_1\cdot\vec{q}_2\,  \frac{\vec{\sigma}_2\cdot\vec{q}_2}{q_2^2 +
  M_\pi^2}\, \int \frac{d^3l}{(2\pi)^3} \frac{l^2}{\omega_l^4}\,,\nn
  \rho_{\rm c7} & = & 
- e\frac{\mathring{g}_A^4
}{8F^4} \ \tau_2^3 \, \frac{\vec{\sigma}_2\cdot\vec{q}_2}{q_2^2 + M_\pi^2}
\int\frac{d^3l}{(2\pi)^3} \left(\vec{\sigma}_1\cdot\vec{l}\,
  \vec{q}_2\cdot\vec{l} - \vec{\sigma}_1\cdot\vec{k}\,
  \vec{q}_2\cdot\vec{k} \right)\frac{1}{\omega_+^2\omega_-^2}\,,
\end{eqnarray}
where 
\beq
\omega_{\pm}^2 =  \left(\vec l \pm \vec k \right)^2 + 4M_\pi^2\, , \quad \quad\omega_{l}^2 =
\vec l \, ^2 + M_\pi^2\,.
\eeq
The class-$3,4,6$ contributions to the current density and 
class-$1,2,3,4,5,8,9$ contributions to the charge density are found to
vanish.

\subsection{Tree level contributions}
\def\theequation{\arabic{section}.\arabic{equation}}
\label{sec5}

\begin{figure}[tb]
\vskip 1 true cm
  \psfrag{plus}{\ $+$}
  \psfrag{arrow}{\ $\rightarrow$}
  \includegraphics[width=0.86\textwidth,keepaspectratio,angle=0,clip]{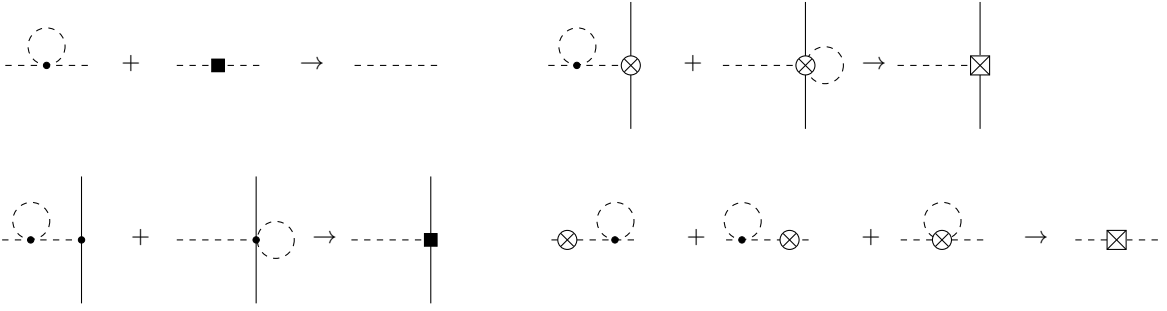}
  \caption{Schematic representation of the renormalization of the pion field
    and the operators $H_{21}^{(3)}$, $J_{21}^{(2)}$ and $J_{02}^{(1)}$  }
  \label{fig:renormalizedfields}
\end{figure}

The loop contributions considered in the previous section do not involve
the ones emerging from pion tadpole diagrams. These must be explicitly taken into
account if one wants to use the values of the renormalized LECs such as $d_i$
determined from e.g.~the pion-nucleon system.  The treatment of the pion
tadpoles in the method of unitary transformation is discussed in detail in
Ref.~\cite{Epelbaum:2002gb}. The pion tadpole contributions emerge from
contractions of the pion field operators when performing the normal ordering
of the effective pion-nucleon Hamiltonian and simply lead to additional vertex
corrections.  
Following Ref.~\cite{Epelbaum:2002gb}, we work with renormalized
pion field and mass defined according to 
\beq
\pi_a^r = Z_\pi^{-1/2} \pi_a\,, \quad \quad Z_\pi = 1 + \delta Z_\pi \,, \quad
\quad M_\pi^2 = M^2 + \delta M_\pi^2\,,   
\eeq
where $a$ denotes the isospin quantum number and $\delta Z_\pi, \; \delta
M_\pi^2/M_\pi^2 \sim \mathcal{O} (Q^2/\Lambda^2)$. At the leading loop order,
$\delta Z_\pi$ and $\delta M_\pi^2$ are given by \cite{Epelbaum:2002gb}
\begin{eqnarray}
\label{eq:zpi}
  \delta Z_\pi & =&  -\frac{2l_4M_\pi^2}{F^2} - \frac{1-10\xi}{F^2}\Delta_\pi\,,\nn
\nonumber
 M_\pi^2 & =&  M^2 \left(1+\frac{2l_3M_\pi^2}{F^2} +
   \frac{1-8\xi}{2F^2}\Delta_\pi \right)\,,
\end{eqnarray}
where the quantity $\Delta_\pi$ is defined in
Eq.~(\ref{def_delta_pi}). 
Notice that in Ref.~\cite{Epelbaum:2002gb} we used the parametrization of the
matrix $U$ with $\xi =0$. We further emphasize that there are no pion
self-energy diagrams since we work with renormalized pion
fields. All effects due to pion self-energy and/or tadpoles are taken into
account by vertex corrections in the normal-ordered effective Hamiltonian. 
This is schematically visualized in Fig.~\ref{fig:renormalizedfields}. 
More precisely, replacing $\pi_a \to
\pi_a^r$ and $M^2 \to M_\pi^2$ in $\mathcal{L}_{\pi\pi}^{(2)}$ and
$\mathcal{L}_{\pi N}^{(1)}$ generates corrections to
$\mathcal{L}_{\pi\pi}^{(4)}$
and $\mathcal{L}_{\pi N}^{(3)}$ (and, of course, in the corresponding Hamilton
densities) driven by $\delta Z_\pi$ and $\delta M_\pi^2$. Further
corrections, $\delta_{\rm NO}$, to the operators $H_{21}^{(3)}$, $J_{21}^{(2)}$
 and $J_{02}^{(1)}$ emerge from taking normal ordering on the  operators
 $H_{23}^{(3)}$, $J_{23}^{(3)}$ and $J_{04}^{(1)}$    
Together with the the wave function
 renormalization of the pion, we obtain the following shifts:
 \begin{eqnarray}
   H_{21}^{(3)} & \rightarrow & H_{21}^{(3)} + H_{21}^{(1)} \left(
     \frac{1}{2}\delta Z_\pi + \delta_{\rm NO}\right) 
   = H_{21}^{(3)} - \, H_{21}^{(1)} \frac{l_4M_\pi^2}{F^2}
   \,, \nn
   {\vec{J}_{21}\,}^{\, (2)} & \rightarrow & {\vec{J}_{21}\,}^{\, (2)} +
   {\vec{J}_{21}\,}^{\, (0)}  \left(
     \frac{1}{2}\delta Z_\pi + \delta_{\rm NO} \right) 
   = {\vec{J}_{21}\,}^{\, (2)} - {\vec{J}_{21}\,}^{\, (0)}
   \left(\frac{l_4M_\pi^2}{F^2} + \frac{1}{2F^2}\Delta_\pi \right)  
\,,\nn  
   {\vec{J}_{02}\,}^{\, (1)} & \rightarrow & {\vec{J}_{02}\,}^{\, (1)} +
   {\vec{J}_{02}\,}^{\, (-1)}
   \left( \delta Z_\pi + \delta_{\rm NO} \right) 
   = {\vec{J}_{02}\,}^{\, (1)} - {\vec{J}_{02}\,}^{\, (-1)} \left(
     \frac{2l_4M_\pi^2}{F^2} + \frac{1}{F^2}\Delta_\pi  \right)\,.
 \end{eqnarray}
 We point out that, as expected, none of the renormalized operators depends on the (arbitrary) value of
 $\xi$.

\begin{figure}[tb]
\vskip 1 true cm
  \includegraphics[width=0.86\textwidth,keepaspectratio,angle=0,clip]{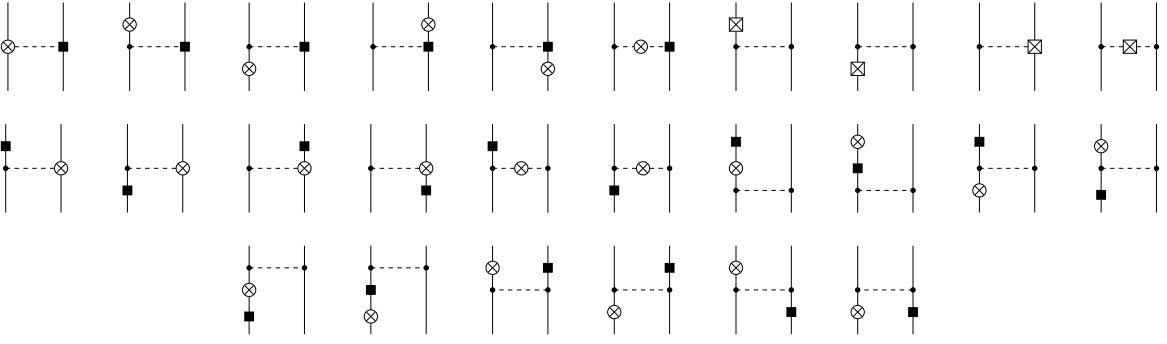}
  \caption{Contributions of the counter terms.  Solid dots are the lowest-order vertices from the effective
    Lagrangian while the crosses represent insertions of the electromagnetic
    vertices as explained in the text. For remaining notation see Fig.~\ref{fig:loops}. }
  \label{fig:counterterms}
\end{figure}

After these preliminary remarks, we are now in the position to discuss the
tree-level contributions to the one-pion exchange current and charge
densities.  The formal operator structure is given by
\begin{eqnarray}
\label{tree_level}
  J_{\rm tree} & = & \eta \biggr[
              - H_{21}^{(3)} \frac{\lambda^1}{E_\pi} J_{21}^{(0)} 
              + H_{21}^{(3)} \frac{\lambda^1}{E_\pi} J_{20}^{(-1)} \frac{\lambda^1}{E_\pi} H_{21}^{(1)} 
              - \frac{1}{2}  J_{20}^{(-1)} \eta H_{21}^{(1)} \frac{\lambda^1}{E_\pi^2} H_{21}^{(3)} 
              - \frac{1}{2}  J_{20}^{(-1)} \eta H_{21}^{(3)} \frac{\lambda^1}{E_\pi^2} H_{21}^{(1)} 
\nn &&{}
              + H_{21}^{(1)} \frac{\lambda^1}{E_\pi^2} H_{21}^{(3)} \eta J_{20}^{(-1)}
              - H_{21}^{(3)} \frac{\lambda^1}{E_\pi^2} H_{21}^{(1)} \eta J_{20}^{(-1)} 
              + H_{21}^{(3)} \frac{\lambda^1}{E_\pi} J_{02}^{(-1)} \frac{\lambda^1}{E_\pi} H_{21}^{(1)} 
              + J_{02}^{(-1)} \frac{\lambda^2}{E_\pi} H_{21}^{(1)}
              \frac{\lambda^1}{E_\pi} H_{21}^{(3)} 
\nn &&{}
              + J_{02}^{(-1)} \frac{\lambda^2}{E_\pi} H_{21}^{(3)}\frac{\lambda^1}{E_\pi} H_{21}^{(1)} 
              + H_{21}^{(1)} \frac{\lambda^1}{E_\pi} J_{20}^{(1)} \frac{\lambda^1}{E_\pi} H_{21}^{(1)}
              - H_{21}^{(1)} \frac{\lambda^1}{E_\pi^2} H_{21}^{(1)} \eta J_{20}^{(1)} 
              - H_{21}^{(1)} \frac{\lambda^1}{E_\pi} J_{21}^{(2)} 
              + H_{21}^{(1)} \frac{\lambda^1}{E_\pi} H_{21}^{(1)} \frac{\lambda^2}{E_\pi} J_{20}^{(1)}
\nn &&{}
              + \frac{1}{2} H_{21}^{(1)} \frac{\lambda^1}{E_\pi} J_{20}^{(1)} \frac{\lambda^1}{E_\pi} H_{21}^{(1)}
           \biggl]\eta+ \textrm{h.c.}\,.
\end{eqnarray}
These operators give rise to diagrams shown in Fig.~\ref{fig:counterterms}.
The explicit form of all vertices entering this expression can be found in
appendix \ref{append_Ham}. 
Evaluating the corresponding matrix elements we obtain the following
expressions for the current density
\begin{eqnarray}
      \vec{J}_{\rm tree} & =  & 2e\frac{\mathring{g}_A\, i}{F^2}\left(d_8
   \tau_2^3 + d_9 \left( \vec{\tau}_1 \cdot\vec{\tau}_2\right) \right)
 \frac{\vec{\sigma}_2\cdot\vec{q}_2}{q_2^2+M_\pi^2}\left[\vec{q}_1
   \times\vec{q}_2\right] -e\, \frac{\mathring{g}_A \, i}{4F^2} \left[
      \vec{\tau}_1\times \vec{\tau}_2\right]^3\,\frac{\vec{\sigma}_2\cdot\vec{q}_2}{q_2^2+M_\pi^2}\,
    \biggl\{2 d_{21} \, 
  \vec{k}\times\left[\vec{q}_2\times\vec{\sigma}_1\right]\nn
&&{}+d_{22}
 \, \vec{k}\times\left[\vec{q}_1\times\vec{\sigma}_1\right]
+     \vec{\sigma}_1\,  \left[2M_\pi^2\, \left(4d_{16} - 2d_{18} - \frac{l_4
           \mathring{g}_A}{F^2}\right) - \frac{\mathring{g}_A}{2}\Delta_\pi\right]
\nn
&&{}- \vec{q}_1
     \frac{\vec{\sigma}_1\cdot\vec{q}_1}{q_1^2 + M_\pi^2} 
 \left[2M_\pi^2\, \left(4d_{16} - 2d_{18} - \frac{l_4
           \mathring{g}_A}{F^2}\right) - \mathring{g}_A \Delta_\pi +
       \mathring{g}_A \, k^2 \frac{l_6}{F^2}   - \mathring{g}_A\,
       \frac{l_6}{F^2} \left(q_1^2 - q_2^2 \right)\right]\biggl\}\, ,
\end{eqnarray}
while the contributions to the charge density vanish. 
This is consistent with the fact that the loop contributions to
the charge density do not contain logarithmic ultraviolet divergences.

\subsection{Renormalization}
\def\theequation{\arabic{section}.\arabic{equation}}
\label{sec6}

\begin{figure}[tb]
\vskip 1 true cm
  \includegraphics[width=0.3\textwidth,keepaspectratio,angle=0,clip]{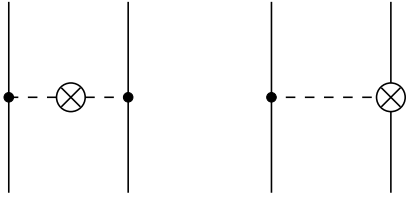}
  \caption{Lowest-order contributions to the pion exchange current operator: the pion-in-flight and 
  seagull graphs. For notation see Fig.~\ref{fig:loops}. }
  \label{fig:leadingorder}
\end{figure}

The expressions given in the previous sections are written in terms of bare
parameters and contain ultraviolet-divergent pieces. These divergences are
cancelled after expressing the bare parameters $M$, $\mathring{g}_A$, $F$, $l_i$ 
and $d_i$ in terms of the corresponding renormalized quantities. When carrying
out renormalization, one should also take into account the contribution
induced by the leading-order ($\mathcal{O} \left( eQ^{-1} \right)$) 
one-pion exchange current shown in 
Fig.~\ref{fig:leadingorder}
\begin{eqnarray}
  \label{eq:LO-Cont}
   \vec{J}^{\; \left( eQ^{-1} \right)}_{1\pi} = e\, \frac{i \mathring{g}_A^2}{4F^2}\left[
     \vec{\tau}_1 \times \vec{\tau}_2\right]^3\,
   \frac{\vec{\sigma}_2\cdot\vec{q}_2}{q_2^2 + M_\pi^2}\, \left( \vec{q}_1
          \frac{\vec{\sigma}_1\cdot\vec{q}_1}{q_1^2 + M_\pi^2}  -
     \vec{\sigma}_1\, \right)\, ,  
 \end{eqnarray}
when expressing the ratio $\mathring{g}_A/F$ in terms of the physical LECs
$g_A/F_\pi$. 
The chiral expansion of this ratio has the form 
\beq
 \frac{g_A}{F_\pi} =  \frac{\mathring{g}_A}{F}\left(1
 -\frac{2g_A^2}{F_\pi^2}\Delta_\pi -
  \frac{M_\pi^2}{F_\pi^2}l_4 + 4\frac{M_\pi^2}{g_A}d_{16}\
  \right)\,,
\eeq
Clearly, this relation holds modulo higher-order corrections. 
The resulting induced  correction at order $\mathcal{O} \left( eQ \right)$ reads: 
\begin{eqnarray}
  \label{eq:Cont_LO_to_NLO}
  \vec{J}^{\; \left( eQ\right) }_{1\pi} = e\frac{g_A^2\,
    i}{2F_\pi^2}\,\frac{\vec{\sigma}_2\cdot\vec{q}_2}{q_2^2 + M_\pi^2}\,\left[
     \vec{\tau}_1 \times \vec{\tau}_2\right]^3\, \left[ \vec{q}_1
     \frac{\vec{\sigma}_1\cdot\vec{q}_1}{q_1^2 + M_\pi^2}
       -
     \vec{\sigma}_1\right]\left(\frac{2 g_A^2}{F_\pi^2} \Delta_\pi +
  \frac{M_\pi^2}{F_\pi^2}l_4 - 4\frac{M_\pi^2}{g_A}d_{16} \right)  \, .
\end{eqnarray}
Notice that at the order considered, one can safely replace $\mathring{g}_A$ and $F$ by the
corresponding renormalized quantities in all expressions given in 
sections~\ref{sec4} and~\ref{sec5}. 

Consider now the LECs $l_i$ and $d_i$ which can be decomposed into the
divergent parts and finite pieces as follows:
 \beqa
  l_i  &=&  l_i^r (\mu) + \gamma_i L = \frac{1}{16\pi^2}\bar{l}_i + \gamma_i L +
  \gamma_i \frac{1}{16\pi^2}\log\left(\frac{M_\pi}{\mu}\right)\,, \nn
  d_i & = & d_i^r (\mu) + \frac{\beta_i}{F^2} L = \bar{d}_i  +
  \frac{\beta_i}{F^2} L + \frac{\beta_i}{16 \pi^2 F^2}
  \log\left(\frac{M_\pi}{\mu}\right)\,,
\eeqa
where the divergent quantity $L$ is defined in Eq.~(\ref{def_L}). 
The corresponding coefficients $\beta_i$ and $\gamma_i$ in the framework of
dimensional regularization (DR) are well known 
\cite{Gasser:1983yg,Ecker:1995rk,Fettes:1998ud,Gasser:2002am}
and read:
\beq
 \beta_8  =  \beta_9 = \beta_{18} = \beta_{22}  = 0, \quad \quad
 \beta_{16} = \frac{1}{2}g_A + g_A^3,\quad  \quad \beta_{21} = -g_A^3,\quad \quad
\gamma_4 =  2, \quad \quad \gamma_6 = - \frac{1}{3}\,.
\eeq
The expressions for all loop integrals that enter the calculation in DR can be
found in Appendix~\ref{app2}. The only exception is the part of the class-7
current proportional to the constant $\delta $, for which we did not succeed to
find a closed expression. 
Inserting the DR expressions for the
integrals entering Eqs.~(\ref{current_dens}), (\ref{charge_dens}) and the pion
tadpole contributions discussed above and replacing the bare LECs in terms of
renormalized ones, one observes that indeed almost all divergences cancel.  
The only remaining divergent part of the current reads 
\begin{eqnarray}
  \vec{J}_{\rm div} & = & -e\frac{g_A^2 \,
  i}{12F_\pi^4}\left[\vec{\tau}_1\times\vec{\tau}_2\right]^3
\frac{\vec{\sigma}_2\cdot\vec{q}_2}{q_2^2 + M_\pi^2} \, \vec{k} \ L\ \left[ \vec{\sigma}_1
\cdot\vec{q}_1  \, \left( 1 - \bar\beta_1\right) \, + g_A^2 \
\vec{\sigma}_1\cdot\vec{q}_2  \left(-2 + 2\beta + \delta \right)\right]\,.
\end{eqnarray}
This implies that we have  to choose $\bar\beta_1 = 1$ and $-2+2\beta + \delta =
0$ in order to be able to  renormalize the current operator. Here and in what
follows, we adopt the choice $\delta  = 0$ and $\beta=1$.

\subsection{Relativistic corrections}
\def\theequation{\arabic{section}.\arabic{equation}}
\label{sec7}

Last but not least, we now discuss the leading relativistic corrections. These emerge from 
the operators in Eq.~(\ref{tree_level}) with the vertices $H_{21}^{(3)}$, $J_{20}^{(1)}$
and $J_{21}^{(2)}$ being replaced by the corresponding relativistic corrections
$\tilde H_{21}^{(3)}$, $\tilde J_{20}^{(1)}$ and $\tilde J_{21}^{(2)}$, respectively, whose explicit form 
is given in appendix \ref{append_Ham}.  In addition, there are contributions emerging 
from insertions of the kinetic energy of the nucleon $\tilde H_{20}^{(2)}$ which
have the form
\begin{eqnarray}
\label{1/mop}
J_{1/m_N} & = & \eta \, \biggr[
            H_{21}^{(1)} \frac{\lambda^1}{E_\pi} \tilde H_{20}^{(2)} \frac{\lambda^1}{E_\pi} J_{21}^{(0)} 
          -   \tilde H_{20}^{(2)} \eta H_{21}^{(1)} \frac{\lambda^1}{E_\pi^2} J_{21}^{(0)} 
          + \bar\beta_7 \, \left(
            H_{21}^{(1)} \frac{\lambda^1}{E_\pi^3} J_{21}^{(0)} \eta  \tilde H_{20}^{(2)} 
          -  \tilde H_{20}^{(2)} \eta H_{21}^{(1)} \frac{\lambda^1}{E_\pi^3} J_{21}^{(0)} 
          \right)\,, \nn
&+&
            H_{21}^{(1)} \frac{\lambda^1}{E_\pi} J_{02}^{(-1)} \frac{\lambda^1}{E_\pi^2} H_{21}^{(1)} \eta  \tilde H_{20}^{(2)} 
          -  H_{21}^{(1)} \frac{\lambda^1}{E_\pi} H_{21}^{(1)} \frac{\lambda^2}{E_\pi}  \tilde H_{20}^{(2)} \frac{\lambda^2}{E_\pi} J_{02}^{(-1)}
          -  H_{21}^{(1)} \frac{\lambda^1}{E_\pi}  \tilde H_{20}^{(2)} \frac{\lambda^1}{E_\pi} H_{21}^{(1)} \frac{\lambda^2}{E_\pi} J_{02}^{(-1)}\nn
          &-&  H_{21}^{(1)} \frac{\lambda^1}{E_\pi}  \tilde H_{20}^{(2)} \frac{\lambda^1}{E_\pi} J_{02}^{(-1)} \frac{\lambda^1}{E_\pi} H_{21}^{(1)}
          +   \tilde H_{20}^{(2)} \eta H_{21}^{(1)} \frac{\lambda^1}{E_\pi^2} H_{21}^{(1)} \frac{\lambda^2}{E_\pi} J_{02}^{(-1)} 
          +   \tilde H_{20}^{(2)} \eta H_{21}^{(1)} \frac{\lambda^1}{E_\pi} H_{21}^{(1)} \frac{\lambda^2}{E_\pi^2} J_{02}^{(-1)} 
           \nn
&+& \bar\beta_4\, \biggr(
             \tilde H_{20}^{(2)} \eta H_{21}^{(1)} \frac{\lambda^1}{E_\pi} H_{21}^{(1)} \frac{\lambda^2}{E_\pi^2} J_{02}^{(-1)} 
          -   \tilde H_{20}^{(2)} \eta J_{02}^{(-1)} \frac{\lambda^2}{E_\pi^2} H_{21}^{(1)} \frac{\lambda^1}{E_\pi} H_{21}^{(1)} 
          \biggl)  + \bar\beta_5\,  \biggr(
             \tilde H_{20}^{(2)} \eta H_{21}^{(1)} \frac{\lambda^1}{E_\pi^2}
            H_{21}^{(1)} \frac{\lambda^2}{E_\pi} J_{02}^{(-1)} \nn
&-&   \tilde H_{20}^{(2)} \eta J_{02}^{(-1)} \frac{\lambda^2}{E_\pi} H_{21}^{(1)} \frac{\lambda^1}{E_\pi^2} H_{21}^{(1)} 
          \biggl) + \bar\beta_6 \, \biggr(
            H_{21}^{(1)} \frac{\lambda^1}{E_\pi} J_{02}^{(-1)} \frac{\lambda^1}{E_\pi^2} H_{21}^{(1)} \eta  \tilde H_{20}^{(2)} 
          -  H_{21}^{(1)} \frac{\lambda^1}{E_\pi^2} J_{02}^{(-1)} \frac{\lambda^1}{E_\pi} H_{21}^{(1)} \eta  \tilde H_{20}^{(2)} 
          \biggl) \nn
&+&
            H_{21}^{(1)} \frac{\lambda^1}{E_\pi} J_{20}^{(-1)} \frac{\lambda^1}{E_\pi^2} H_{21}^{(1)} \eta  \tilde  H_{20}^{(2)} 
          -  H_{21}^{(1)} \frac{\lambda^1}{E_\pi}  \tilde H_{20}^{(2)} \frac{\lambda^1}{E_\pi} J_{20}^{(-1)} \frac{\lambda^1}{E_\pi} H_{21}^{(1)} 
          - \frac{1}{2}  H_{21}^{(1)} \frac{\lambda^1}{E_\pi^3} H_{21}^{(1)} \eta  \tilde H_{20}^{(2)} \eta J_{20}^{(-1)} \nn
          &+& \frac{1}{2}  H_{21}^{(1)} \frac{\lambda^1}{E_\pi^2}  \tilde H_{20}^{(2)} \frac{\lambda^1}{E_\pi} H_{21}^{(1)} \eta J_{20}^{(-1)} 
          + \frac{1}{2}  H_{21}^{(1)} \frac{\lambda^1}{E_\pi}  \tilde H_{20}^{(2)} \frac{\lambda^1}{E_\pi^2} H_{21}^{(1)} \eta J_{20}^{(-1)} 
          - \frac{1}{2}  \tilde  H_{20}^{(2)} \eta H_{21}^{(1)} \frac{\lambda^1}{E_\pi^3} H_{21}^{(1)} \eta J_{20}^{(-1)} 
          \nn
& +& \bar\beta_2\, \eta \, \biggr(
            H_{21}^{(1)} \frac{\lambda^1}{E_\pi^2} J_{20}^{(-1)} \frac{\lambda^1}{E_\pi} H_{21}^{(1)} \eta  \tilde H_{20}^{(2)} 
          -  H_{21}^{(1)} \frac{\lambda^1}{E_\pi} J_{20}^{(-1)} \frac{\lambda^1}{E_\pi^2} H_{21}^{(1)} \eta  \tilde H_{20}^{(2)} 
          \biggl) +  \bar\beta_3\,  \biggr(
           \tilde  H_{20}^{(2)} \eta H_{21}^{(1)} \frac{\lambda^1}{E_\pi^3}
           H_{21}^{(1)} \eta J_{20}^{(-1)} \nn
&-&  \tilde H_{20}^{(2)} \eta J_{20}^{(-1)} \eta H_{21}^{(1)} \frac{\lambda^1}{E_\pi^3} H_{21}^{(1)} 
          \biggl) \biggl] \eta + \textrm{h.c.}\,,
\end{eqnarray}
where the constants $\bar\beta_i$ are defined in Eq.~(\ref{betas}).  
The additional $\eta$-space UTs considered so
far did not involve $1/m_N$-corrections. The unitary ambiguity of the
leading relativistic corrections can be parametrized in terms of the following
two additional UTs: 
\begin{eqnarray}
  U^\prime & = & e^{S^\prime}\, , \nn
  S^\prime & = & \bar\beta_8 S_8 + \bar\beta_9 S_9\,, \nn
\end{eqnarray}
with two new constants $\bar\beta_8$ and $\bar\beta_9$  and the operators $S_{8,9}$ given by   
\begin{eqnarray}
  S_8 & = & \eta \left[ \tilde H_{20}^{(2)} \eta H_{21}^{(1)}\frac{\lambda_1}{E_\pi^3}
  H_{21}^{(1)} - H_{21}\frac{\lambda_1}{E_\pi^3}
  H_{21}^{(1)} \eta \tilde H_{20}^{(2)} \right]\eta\,, \nn
S_9 & = & \eta \left[ \tilde H_{21}^{(3)} \frac{\lambda^1}{E_\pi^2}H_{21}^{(1)} -
  H_{21}^{(1)} \frac{\lambda^1}{E_\pi^2} \tilde H_{21}^{(3)}\right] \eta\,.
\end{eqnarray}
Notice that the operator $S_9$ with $\tilde H_{21}^{(3)}$ being replaced by 
$H_{21}^{(3)}$ vanishes which is why the corresponding UT was not considered 
in section \ref{sec5}.
The effects of these UTs in connection with the nuclear
potentials and currents have already been investigated, see
\cite{Friar:1999sj,Adam:1993zz} and references therein. 
In particular, these UTs affect $1/m_N^2$-corrections to the one-pion
exchange and $1/m_N$-corrections to the two-pion exchange nucleon-nucleon
potentials which appear at N$^3$LO in the chiral expansion. 
The form of the relativistic corrections adopted in the N$^3$LO potential of
Ref.~\cite{Epelbaum:2004fk} corresponds to the choice $\bar\beta_8 = 1/4$ and
$\bar\beta_9=0$.  The UTs driven by $S_8$ and $S_9$ also induce
additional contributions to the current operator given by 
\beqa
\label{1/mind}
  \delta J_{\rm 1/m_N} &=&  \eta\biggl[ \bar\beta_8 \bigg( H_{21}^{(1)} \frac{\lambda^1}{E_\pi^3}
  H_{21}^{(1)} \eta \tilde{H}_{20}^{(2)} \eta J_{20}^{(-1)} - \tilde{H}_{20}^{(2)} \eta
  H_{21}^{(1)} \frac{\lambda^1}{E_\pi^3} H_{21}^{(1)} \eta J_{20}^{(-1)}
  \bigg)   + \bar\beta_9 \, \bigg( H_{21}^{(1)} \frac{\lambda^1}{E_\pi^2}
  \tilde{H}_{21}^{(3)} \eta J_{20}^{(-1)}  \nn
&-& \tilde{H}_{21}^{(3)} \frac{\lambda^1}{E_\pi^2}
  H_{21}^{(1)} \eta J_{20}^{(-1)} \bigg) \biggr] \eta+ \textrm{h.c.}\,. 
\eeqa
Evaluating matrix elements of the operators given in Eqs.~(\ref{1/mop}) and
(\ref{1/mind}) we find no contributions to the current density. For the charge
density we obtain the following result:
\begin{eqnarray}
\rho_{1/m_N} 
 & = & \frac{e g_A^2}{16F_\pi^2m_N} \frac{1}{q_2^2+M_\pi^2}\biggl\{(1-2\bar\beta_9) \left(\tau_2^3 +
   \vec{\tau}_1\cdot\vec{\tau}_2 \right) \vec{\sigma}_1\cdot\vec{k}
 \vec{\sigma}_2\cdot\vec{q}_2-  i \left[
   \vec{\tau}_1\times\vec{\tau}_2\right]^3 \biggl[(1+2\bar\beta_9) \left(\vec{\sigma}_1\cdot\vec{k}_1
 \vec{\sigma}_2\cdot\vec{q}_2 - \vec{\sigma}_2\cdot\vec{k}_2
 \vec{\sigma}_1\cdot\vec{q}_2 \right)
\nn
&- & 2\, 
\frac{\vec{\sigma}_1\cdot\vec{q}_1}{q_1^2+M_\pi^2}\vec{\sigma}_2\cdot\vec{q}_2\, 
  \vec{q}_1\cdot\vec{k}_1  \biggr]\biggr\}+ \frac{i e g_A^2\,
        }{16 F_\pi^2 m_N } \frac{\vec{\sigma}_2\cdot\vec{q}_2
      }{(q_2^2+M_\pi^2)^2} \,\vec{\sigma}_1\cdot\vec{q}_2
\, \biggl[(1-2\bar\beta_8)\,i (\tau_2^3 + \left(
  \vec{\tau_1}\cdot\vec{\tau}_2\right)) \, \vec{q}_2\cdot\vec{k} \nn
&+&  \left[
  \vec{\tau}_1 \times \vec{\tau}_2\right]^3  
\biggl(\left(2\bar\beta_3\,
  \vec{q}_1 + 2\bar\beta_8\, \vec{q}_2 -  \vec{q}_2\right)\cdot\vec{k}_1 
+  \left(2\bar\beta_3\,
   - 2\bar\beta_8 - 1\right) \vec{q}_2\cdot\vec{k}_2\,\biggr)\biggr]\,.
\end{eqnarray}
Here we have introduced $\vec{k}_{1,2}  =  \vec{p}_{1,2}^{\, \prime} + 
\vec{p}_{1,2}$. In addition to the constants $ \bar\beta_{8,9}$ which parametrize
the $1/m_N$-dependent UTs and also show up in the expressions for the one-pion exchange 
potential, the exchange charge density in the above expression also depends 
on the arbitrary constant $ \bar\beta_3$, which shows up neither in the potential nor in the 
remaining contributions to the exchange charge and current densities. We found that 
the corresponding UT affects the single-nucleon charge operator. Moreover, renormalizability of the 
single-nucleon charge operator enforces the choice $ \bar\beta_3 = 0$.

\subsection{Final results}
\def\theequation{\arabic{section}.\arabic{equation}}
\label{sec8}

In this section we summarize the final, renormalized expressions for the
current and charge densities at order $eQ$, 
\beq
  \vec{J}_{\rm 1\pi}  =  \sum_X \vec{J}_{\rm cX} + \vec{J}_{\rm tree}\,, \quad
  \quad \mbox{and}\quad\quad
 \rho_{\rm 1\pi}  = \sum_X \rho_{\rm cX}  + \rho_{1/m_N}\,.
\eeq
The obtained results in momentum space read: 
\begin{eqnarray}
\label{1pi_current_final}
  \vec{J}_{1\pi} & = &
 \frac{\vec{\sigma}_2\cdot\vec{q}_2}{q_2^2+M_\pi^2}\left[\vec{q}_1
   \times\vec{q}_2\right] \left[
   \tau_2^3 \, f_1(k )+  \vec{\tau}_1 \cdot\vec{\tau}_2\, f_2(k ) \right] +
 \left[\vec{\tau}_1\times\vec{\tau}_2 \right]^3
  \frac{\vec{\sigma}_2\cdot\vec{q}_2}{q_2^2 + M_\pi^2} \biggl\{ 
  \vec{k}\times\left[\vec{q}_2\times\vec{\sigma}_1 
    \right] f_3(k )\nn
 & + &  
  \, \vec{k}\times\left[\vec{q}_1\times\vec{\sigma}_1 \right] f_4(k )+
 \vec{\sigma}_1 \cdot
    \vec{q}_1\, \left( \frac{\vec{k}}{k^2} - \frac{\vec{q}_1
      }{q_1^2+M_\pi^2}  \right) f_5(k) + \,\biggl[\frac{\vec{\sigma}_1\cdot
      \vec{q}_1}{q_1^2+M_\pi^2}\vec{q}_1  -  \vec{\sigma}_1 \biggr]f_6(k )\biggl\} \,,
  \end{eqnarray}
where the scalar functions $f_i(k)$ are given by
\begin{eqnarray}
  f_1\left(k \right) & = & 2i e\frac{g_A}{F_\pi^2}\, \bar{d}_8\, , \nn 
  f_2\left(k \right)  &=&  2i e\frac{g_A}{F_\pi^2}\, \bar{d}_9 \, , \nn  
  f_3\left(k \right)  &=&  - ie\frac{g_A}{64 F_\pi^4\pi^2}
  \left[\,g_A^3\,\left( 2L(k)-1\right)  + 32 F_\pi^2\pi^2 \bar{d}_{21}
  \right]\, , \nn
  f_4\left(k \right) & = & - ie\frac{g_A}{4 F_\pi^2}
  \,\bar{d}_{22} \, , \nn  
  f_5\left(k \right)  &=&  - ie \frac{g_A^2}{384 F_\pi^4\pi^2}\left[2(4M_\pi^2 +
      k^2)L(k) +\left(6 \, \bar{l}_6 -\frac{5}{3}\right)k^2   - 8M_\pi^2
    \right]\, , \nn  
  f_6\left(k \right)  &=&  - ie\frac{g_A}{ F_\pi^2}  M_\pi^2
  \,\bar{d}_{18}  \,,
\end{eqnarray}
and the loop function $L(k)$ is defined in Eq.~(\ref{loop_functions}).  
The one-pion exchange charge density has the following form:
\begin{eqnarray}
\label{1pi_charge_final}
  \rho_{1\pi} & = &
  \frac{\vec{\sigma}_2\cdot\vec{q}_2}{q_2^2 + M_\pi^2}\tau_2^3\biggr[ 
  \vec{\sigma}_1\cdot\vec{k} \, \vec{q}_2 \cdot\vec{k} f_7( k ) +
  \vec{\sigma}_1\cdot\vec{q}_2 f_8( k )\biggl] \;+ \; \frac{e g_A^2}{16F_\pi^2m_N} \frac{1}{q_2^2 + M_\pi^2} \biggl\{(1-2\bar\beta_9) \left(\tau_2^3 +
   \vec{\tau}_1\cdot\vec{\tau}_2 \right) \vec{\sigma}_1\cdot\vec{k}
 \vec{\sigma}_2\cdot\vec{q}_2\nn
  & - & 
  i(1+2\bar\beta_9) \left[
   \vec{\tau}_1\times\vec{\tau}_2\right]^3\biggl[
    \left(\vec{\sigma}_1\cdot\vec{k}_1
 \vec{\sigma}_2\cdot\vec{q}_2 - \vec{\sigma}_2\cdot\vec{k}_2
 \vec{\sigma}_1\cdot\vec{q}_2 \right)- 2\, \frac{\vec{\sigma}_1\cdot\vec{q}_1}{q_1^2+M_\pi^2}\vec{\sigma}_2\cdot\vec{q}_2\, 
  \vec{q}_1\cdot\vec{k}_1  \biggr]\biggr\}  + \frac{e g_A^2\,
        }{16 F_\pi^2 m_N } \frac{ \vec{\sigma}_1\cdot\vec{q}_2  \, \vec{\sigma}_2\cdot\vec{q}_2
     }{(q_2^2+M_\pi^2)^2} \nn
 & \times & \biggl[(2\bar\beta_8-1)(\tau_2^3 + \left(
  \vec{\tau_1}\cdot\vec{\tau}_2\right)) \, \vec{q}_2\cdot\vec{k}  +   i \left[
  \vec{\tau}_1 \times \vec{\tau}_2\right]^3  
\left(\left( 2\bar\beta_8 -  1\right) \vec{q}_2 \cdot\vec{k}_1 
-  \left(
    2\bar\beta_8 + 1\right) \vec{q}_2\cdot\vec{k}_2\,\right)\biggr]\,,
\end{eqnarray}
where we have introduced
\begin{eqnarray}
f_7( k ) & = & e \frac{g_A^4}{64F_\pi^4\pi} \left[A(k) +
    \frac{M_\pi-4M_\pi^2 \,A(k)}{k^2} \right]\, , \nn 
f_8( k ) &=&   e \frac{g_A^4}{64F_\pi^4\pi}  \left[(4M_\pi^2+k^2)A(k) - M_\pi \right] \, .
\end{eqnarray}
The loop function $A(k)$ is defined in Eq.~(\ref{loop_functions}).

\section{Short-range currents}
\def\theequation{\arabic{section}.\arabic{equation}}
\setcounter{equation}{0}
\label{sec10}

\begin{figure}[tb]
\vskip 1 true cm
  \includegraphics[width=0.86\textwidth,keepaspectratio,angle=0,clip]{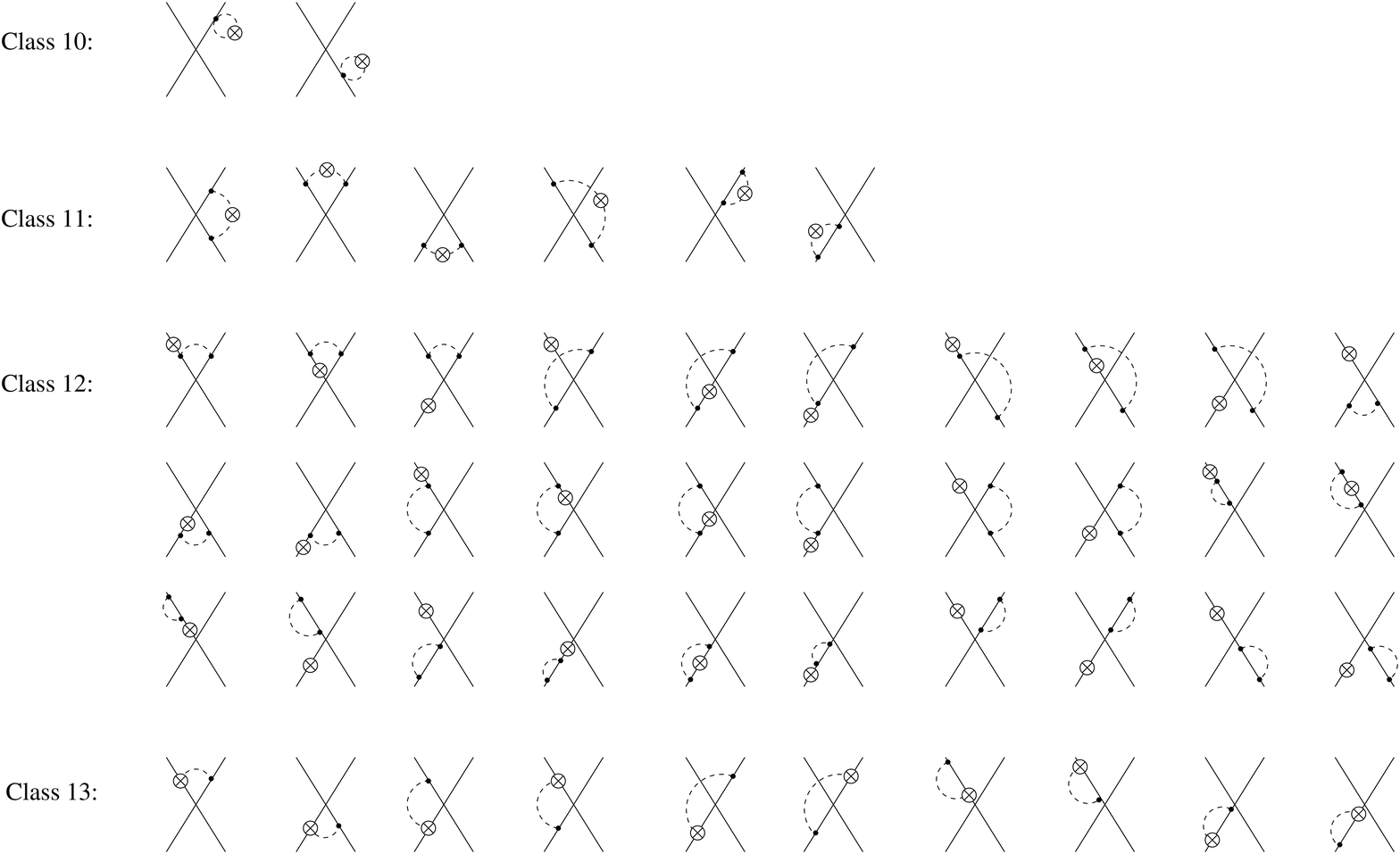}
  \caption{Contributions of the short-range currents.  Solid dots are the
    lowest-order vertices from the effective
    Lagrangian while the circle-crosses represent insertions of the electromagnetic
    vertices as explained in the text. For remaining notation see Fig.~\ref{fig:loops}. }
  \label{fig:contactterms}
\end{figure}

We now consider the short-range contributions. 
The formal structure of the currents involving the leading-order
four-nucleon contact interactions $H_{40}^{(2)}$ can be decomposed into four
classes $10 \ldots 13$ as visualized in Fig.~\ref{fig:contactterms}. Including
the contributions induced by the UTs in Eq.~(\ref{UTbeta}), we obtain the
following algebraic structure: 
\begin{eqnarray}
\label{short_range_induced}
J_{\rm c10} & = &  \eta\biggr[ H_{22}^{(2)} \frac{\lambda^2}{E_\pi} H_{40}^{(2)} \frac{\lambda^2}{E_\pi} J_{02}^{(-1)} 
          -  H_{40}^{(2)} \eta H_{22}^{(2)} \frac{\lambda^2}{E_\pi^2} J_{02}^{(-1)} 
          -\bar\beta_1 \biggr( H_{22}^{(2)} \frac{\lambda^2}{E_\pi} H_{40}^{(2)} \frac{\lambda^2}{E_\pi} J_{02}^{(-1)} 
          -  H_{40}^{(2)} \eta H_{22}^{(2)} \frac{\lambda^2}{E_\pi^2} J_{02}^{(-1)} 
          \biggl) \biggl]\eta + \textrm{h.c.}\,, \nn
J_{\rm c11} & = &
           \eta\biggr[
            H_{40}^{(2)} \eta H_{21}^{(1)} \frac{\lambda^1}{E_\pi^2} J_{02}^{(-1)} \frac{\lambda^1}{E_\pi} H_{21}^{(1)} 
          -  H_{21}^{(1)} \frac{\lambda^1}{E_\pi} J_{02}^{(-1)} \frac{\lambda^1}{E_\pi} H_{40}^{(2)} \frac{\lambda^1}{E_\pi} H_{21}^{(1)} 
          +  J_{02}^{(-1)} \frac{\lambda^2}{E_\pi^2} H_{21}^{(1)} \frac{\lambda^1}{E_\pi} H_{21}^{(1)} \eta H_{40}^{(2)} \nn
          &+&  J_{02}^{(-1)} \frac{\lambda^2}{E_\pi} H_{21}^{(1)} \frac{\lambda^1}{E_\pi^2} H_{21}^{(1)} \eta H_{40}^{(2)} 
          -  J_{02}^{(-1)} \frac{\lambda^2}{E_\pi} H_{21}^{(1)} \frac{\lambda^1}{E_\pi} H_{40}^{(2)} \frac{\lambda^1}{E_\pi} H_{21}^{(1)} 
          -  J_{02}^{(-1)} \frac{\lambda^2}{E_\pi}H_{40}^{(2)}\frac{\lambda^2}{E_\pi} H_{21}^{(1)} \frac{\lambda^1}{E_\pi}H_{21}^{(1)} 
          \nn
& + & \bar\beta_4 
          \biggr(
           H_{40}^{(2)} \eta H_{21}^{(1)} \frac{\lambda^1}{E_\pi} H_{21}^{(1)} \frac{\lambda^2}{E_\pi^2} J_{02}^{(-1)} 
          - H_{40}^{(2)} \eta J_{02}^{(-1)} \frac{\lambda^2}{E_\pi^2} H_{21}^{(1)} \frac{\lambda^1}{E_\pi} H_{21}^{(1)} 
          \biggl) +\bar \beta_5
             \biggr(
            H_{40}^{(2)} \eta H_{21}^{(1)} \frac{\lambda^1}{E_\pi^2}
            H_{21}^{(1)} \frac{\lambda^2}{E_\pi} J_{02}^{(-1)} \nn
& -&  H_{40}^{(2)} \eta J_{02}^{(-1)} \frac{\lambda^2}{E_\pi} H_{21}^{(1)} \frac{\lambda^1}{E_\pi^2} H_{21}^{(1)} 
          \biggl)+\bar \beta_6 
          \biggr(
            H_{40}^{(2)} \eta H_{21}^{(1)} \frac{\lambda^1}{E_\pi^2} J_{02}^{(-1)} \frac{\lambda^1}{E_\pi} H_{21}^{(1)} 
          -  H_{40}^{(2)} \eta H_{21}^{(1)} \frac{\lambda^1}{E_\pi} J_{02}^{(-1)} \frac{\lambda^1}{E_\pi^2} H_{21}^{(1)} \biggl) \biggl]\eta+ \textrm{h.c.}\,, \nn
J_{\rm c12} & = & \eta\biggr[
            H_{21}^{(1)} \frac{\lambda^1}{E_\pi} J_{20}^{(-1)} \frac{\lambda^1}{E_\pi^2} H_{21}^{(1)} \eta H_{40}^{(2)} 
          -  H_{21}^{(1)} \frac{\lambda^1}{E_\pi} H_{40}^{(2)} \frac{\lambda^1}{E_\pi} J_{20}^{(-1)} \frac{\lambda^1}{E_\pi} H_{21}^{(1)}
          - \frac{1}{2}  J_{20}^{(-1)} \eta H_{21}^{(1)} \frac{\lambda^1}{E_\pi^3} H_{21}^{(1)} \eta H_{40}^{(2)} \nn
          &+& \frac{1}{2}  J_{20}^{(-1)} \eta H_{21}^{(1)} \frac{\lambda^1}{E_\pi^2} H_{40}^{(2)} \frac{\lambda^1}{E_\pi} H_{21}^{(1)} 
          + \frac{1}{2}  J_{20}^{(-1)} \eta H_{21}^{(1)} \frac{\lambda^1}{E_\pi} H_{40}^{(2)} \frac{\lambda^1}{E_\pi^2} H_{21}^{(1)} 
          - \frac{1}{2}  J_{20}^{(-1)} \eta H_{40}^{(2)} \eta H_{21}^{(1)}\frac{\lambda^1}{E_\pi^3} H_{21}^{(1)}
           \nn
& + &  
          \bar\beta_2 \biggr(
            H_{40}^{(2)} \eta H_{21}^{(1)} \frac{\lambda^1}{E_\pi}J_{20}^{(-1)} \frac{\lambda^1}{E_\pi^2} H_{21}^{(1)}
          -  H_{40}^{(2)} \eta H_{21}^{(1)} \frac{\lambda^1}{E_\pi^2} J_{20}^{(-1)} \frac{\lambda^1}{E_\pi} H_{21}^{(1)} 
          \biggl) +\bar\beta_3\eta\biggr( 
            H_{40}^{(2)} \eta H_{21}^{(1)} \frac{\lambda^1}{E_\pi^3}
            H_{21}^{(1)} \eta J_{20}^{(-1)}\nn 
          &-&  H_{21}^{(1)} \frac{\lambda^1}{E_\pi^3} H_{21}^{(1)} \eta J_{20}^{(-1)} \eta H_{40}^{(2)}
          \biggl) \biggl]\eta+ \textrm{h.c.}\,, \nn
J_{\rm c13} & = &  \eta\biggr[
             J_{21}^{(0)} \frac{\lambda^1}{E_\pi} H_{40}^{(2)} \frac{\lambda^1}{E_\pi} H_{21}^{(1)}
          -  J_{21}^{(0)} \frac{\lambda^1}{E_\pi^2} H_{21}^{(1)} \eta H_{40}^{(2)} 
          + \bar\beta_7 \, \eta \left(
             H_{21}^{(1)} \frac{\lambda^1}{E_\pi^3} J_{21}^{(0)} \eta H_{40}^{(2)} 
          -  H_{40}^{(2)} \eta H_{21}^{(1)} \frac{\lambda^1}{E_\pi^3} J_{21}^{(0)} 
          \right) \biggl] \eta+ \textrm{h.c.}\,.
\end{eqnarray}
We found that only the class-11 matrix elements yield
non-vanishing contributions to the current density:
\beq
    \vec{J}_{\rm c11} =  - e\frac{
    g_A^2 \, i}{4F_\pi^2}\, C_T \int \frac{d^3l}{(2\pi)^3} \,
  \vec{l} \biggl[2\,\vec{l}\cdot\left[\vec{k}\times\vec{\sigma}_1 \right]\tau_2^3
  +  \vec{\sigma}_1\cdot\vec{l}
    \, \vec{\sigma}_2\cdot\vec{k}\left[\vec{\tau}_1\times\vec{\tau}_2
    \right]^3\biggr] \frac{\omega_+^2 + \omega_+ \omega_- +
      \omega_-^2}{\omega_+^3\omega_-^3(\omega_+ + \omega_-)}\,.
\eeq
We find, however, that the resulting contribution to the current
vanishes after performing antisymmetrization of the two-nucleon
states. 

The  tree contributions emerge  from gauging the subleading contact
interactions in the Lagrangian $\mathcal{L}_{NN}^{(2)}$ and the two 
new gauge-invariant terms proportional to the LECs $L_{1,2}$:  
\begin{eqnarray}
\label{con2}
 \vec{J}_{\rm contact} & = & e\,\frac{i}{16}\left[ \vec{\tau}_1 \times \vec{\tau}_2\right]^3 \,\biggl[\left(C_2 +3C_4 + C_7\right)
    \, \vec{q}_1 - \left(-C_2 + C_4 + C_7 \right) \,
  \, \left(\vec{\sigma}_1 \cdot\vec{\sigma}_2 \right)\, \, \vec{q}_1 
 + C_7 \, \left(  \vec{\sigma}_2\cdot\vec{q}_1
      \,  \vec{\sigma}_1+ \, \vec{\sigma}_1\cdot\vec{q}_1
     \,  \vec{\sigma}_2 \right)\biggr]\nn
 &- &e\,\frac{C_5 \, i}{16}\, \tau_1^3 \,
 \left[\left(\vec{\sigma}_1 + \vec{\sigma}_2 \right)\times\vec{q}_1
   \right] +  i e L_1 \, \tau_1^3 \,
\left[\left(\vec{\sigma}_1 - \vec{\sigma}_2 \right)\times\vec{k}\right] + i e L_2 \, \left[\left(\vec{\sigma}_1 + \vec{\sigma}_2
  \right)\times\vec{q}_1 \right] \, . 
\end{eqnarray}
It is reassuring to note that all the divergences of the two-pion exchange
loop integrals are cancelled by the same redefinition of $C_i$ that is needed to
renormalize the potential~\cite{Epelbaum:2003gr}. From the two LECs that are
genuine to the current operator, only $L_1$ gets renormalized. Employing DR
and the $\overline{\textrm{MS}}$-scheme, the relation between the bare and renormalized
LEC $L_1$ has the form  
\begin{eqnarray}
  L_1 & = & \bar{L}_1 + \frac{g_A^2 -3g_A^4}{8F_\pi^4 } \Delta_\pi \,.
\end{eqnarray}

The charge density is completely given by the class-11,12 loop diagrams:
\begin{eqnarray}
 \rho_{\rm c11} & = & -e\frac{
    g_A^2}{4F_\pi^2}\, C_T \, \tau_1^3 \int \frac{d^3l}{(2\pi)^3}
  \left( \vec{\sigma}_1\cdot\vec{\sigma}_2\, (k^2-l^2)-
    \vec{\sigma}_1\cdot\vec{k}\,\vec{\sigma}_2\cdot\vec{k} + \vec{\sigma}_1\cdot\vec{l}\,\vec{\sigma}_2\cdot\vec{l}
  \,\right)\frac{1}{\omega_+^2 \omega_-^2} 
 \,, \nn
 \rho_{\rm c12} & = & - e \frac{g_A^2 }{3 F_\pi^2}
  \, C_T \,  \, \tau_1^3 \vec{\sigma}_1 \cdot\vec{\sigma}_2 \int
  \frac{d^3l}{(2\pi)^3} \frac{l^2}{\omega_l^4}\,.
\end{eqnarray}

In DR, the integrals entering these expressions are finite, and the 
result for the short-range charge density reads
\begin{equation}
\label{con3}
\rho_{\rm contact} =  C_T\, \tau_1^3\, \left[
      \vec{\sigma}_1\cdot\vec{k}\,\vec{\sigma}_2\cdot\vec{k}\, f_{9}(k ) + \vec{\sigma}_1\cdot\vec{\sigma}_2 \,
    f_{10}( k )\right]  \,,
\end{equation}
where 
\begin{eqnarray}
f_{9}( k ) &=& e \frac{g_A^2}{32F_\pi^2\pi} \left(A(k) +
    \frac{M_\pi-4M_\pi^2 \,A(k)}{k^2} \right)\, , \nn 
f_{10}( k ) &=& e\frac{g_A^2}{32 F_\pi^2\pi}\, \left(M_\pi-(4M_\pi^2 + 3k^2)\,A(k) \right)\,.
\end{eqnarray}
Notice that for antisymmetric nuclear states, the two structures
in Eq.~(\ref{con3}) can be combined into a single one.  
Equations (\ref{con2}) and (\ref{con3}) represent our final
results for the short-range contributions.

\section{Comparison with the work by Pastore et al.}
\def\theequation{\arabic{section}.\arabic{equation}}
\setcounter{equation}{0}
\label{sec_pastore}

We now compare our results given in the previous sections with the ones obtained 
by Pastore et al.~\cite{Pastore:2008ui,Pastore:2009is,Pastore:2011ip}. Below, we list the 
(numerous) differences and, in some cases, comment on their possible origin. 
\begin{itemize}
\item
We begin with the exchange current density given in Eq.~(\ref{1pi_current_final}).
 Our result for the pion loop contributions (i.e.~terms proportional to the loop function $L(k)$) 
agrees with the one of \cite{Pastore:2009is} for the class-9 operator, see 
the contribution $\propto L(k)$ in $f_5$. As Pastore et al., we also
find, that the class-8 operator is cancelled by a part of the class-5
operator. The rest of the class-5 operator is not mentioned 
in~\cite{Pastore:2009is}. As shown in section~\ref{sec6}, this part vanishes in
our treatment due to the renormalizability constraint.  For the seagull current, see 
the contribution $\propto L(k)$ in $f_3$,  we obtain 
a completely different result with even a different isospin dependence: 
$[ \vec \tau_1 \times \vec \tau_2 ]^3$  as compared to $\tau_{1,2}^3$ in \cite{Pastore:2009is}, 
see Eq.~(3.36) in their work.  This should not come as a surprise given the fact that 
Pastore et al.~did not succeed to extract the (truly) irreducible part of the amplitude 
for this particular topology, see the discussion in appendix E of \cite{Pastore:2009is}. 
In particular, they even encountered some non-Hermitian contributions which
then were ignored.
\item
We also disagree on the tree contributions to the current density except  the one $\propto d_{21}$.  
In particular, our terms $\propto d_{8,9}$ have a different sign. Further, we find independent 
contributions from both LECs $d_{21}$ and $d_{22}$, while in  \cite{Pastore:2009is} they only 
appear in a linear combination $2d_{21} + d_{22}$.
Finally, Pastore et al.~miss the contributions
from the LEC $d_{18}$ (which accounts for the Goldberger-Treiman discrepancy) and   
$l_6$. Last but not least, we emphasize that our result for the current operator in  
Eq.~(\ref{1pi_current_final}) depends on renormalized LECs $\bar d_{8,9,18,21,22}$ and $\bar l_6$
which, of course,  can be taken from other reactions such as e.g.~pion-nucleon
scattering \cite{Fettes:2001cr}.
\item
We now turn to the one-pion exchange charge density operator given in Eq.~(\ref{1pi_charge_final}).
The expressions for the leading relativistic corrections $\propto 1/m_N$ are, of course, 
not new and agree with the ones given in \cite{Pastore:2011ip}.\footnote{We provide a somewhat 
more general result than the one of \cite{Pastore:2011ip} by including effects due to both UTs available 
at this order (terms proportional to $\bar \beta_{8,9}$). The choice of these parameters 
consistent with the two-nucleon potentials  of Ref.~\cite{Epelbaum:2004fk} corresponds to 
$\bar \beta_8 = 1/4$ and $\bar \beta_9=0$. Contributions proportional to $\bar \beta_9$ are not 
considered in Ref.~\cite{Pastore:2011ip}.}  We, however, also obtain nonvanishing pion loop 
contributions to the exchange charge density, see the $f_{7,8}$ terms in  Eq.~(\ref{1pi_charge_final}), 
which are not considered in Ref.~\cite{Pastore:2011ip}.
\item
Finally, our expressions for the pion loop contribution to the short-range current and charge 
operators  also strongly disagree with the ones given in 
Refs.~\cite{Pastore:2009is} and \cite{Pastore:2011ip}, respectively.  In particular, the results obtained
by Pastore et al. depend on both leading-order LECs $C_S$ and $C_T$, while there is no dependence
on $C_S$ in our case. Morover, we find that the short-range pion loop
contribution to the current density vanishes completely upon performing
antisymmetrization.  The origin of these discrepancies might be related to the unitary ambiguity 
of the nuclear potential and current operators. As discussed in the previous sections, we include 
in our derivation a large number of additional UTs which are possible at the given order in the chiral 
expansion.  In particular, pion loop contributions to short-range current/charge operators are 
affected by the strong UT defined in Eq.~(3.48) of Ref.~\cite{Epelbaum:2007us} 
and $\mathcal{A_\mu}$-dependent 
UTs which induce  additional operators listed in Eq.~(\ref{short_range_induced}).
As a consequence, the resulting short-range currents might be expected to be strongly 
scheme dependent (i.e.~dependent on the a priori unknown 
angles of these additional UTs). It is 
\emph{the renormalizability requirement} of the nuclear potentials and currents that 
provides strong constraints on the choice of the additional UTs, see the detailed  discussion 
in Refs.~\cite{Kolling:2009iq,Epelbaum:2007us} and in section \ref{sec1},  
and leads finally to unambiguous expressions for the (static) nuclear potentials and 
current/charge operators at the considered order.  The observed differences for the 
short-range operators suggest that the results of Ref.~\cite{Pastore:2011ip} might 
correspond to a different choice of the additional UTs as compared to the one 
adopted in our work.\footnote{For example, we could easily generate terms proportional
to $C_S$ by choosing additional UTs in a different way.} Our findings, however, imply
that such a different choice would result in impossibility to obtain renormalized expressions 
for the nuclear forces and/or the current operator. 
\end{itemize}

\section{Summary}
\def\theequation{\arabic{section}.\arabic{equation}}
\setcounter{equation}{0}
\label{sec_summ}

The results of our work can be summarized as follows: 
\begin{itemize}
\item
We applied the method of unitary transformation to work out the leading loop
contributions to the one-pion exchange and short-range two-nucleon
electromagnetic current and charge densities. The \emph{renormalized} expressions for 
the one-pion exchange charge and current operators are given for the first time.  
\item
We discuss in detail renormalization of the one-pion exchange contributions
which provides a stringent test of our theoretical approach. More precisely,
\emph{all} emerging ultraviolet divergences have to be absorbed into
redefinition of the low-energy constants $l_i$ and $d_i$ entering the Lagrangians  
$\Lag_{\pi\pi}^{(4)}$ and $\Lag_{\pi N}^{(3)}$, respectively. There is no
freedom in this procedure as the corresponding $\beta$-functions of all these
LECs in DR are fixed  and well known. We demonstrate, that it is indeed
possible to renormalize the one-loop contributions provided one makes use of
the freedom to employ additional unitary transformations.  
\item
We succeeded to obtain compact, analytical expressions for the current and charge densities both
in momentum and coordinate spaces which can be used in future numerical calculations. 
\item
Finally, we provide a detailed comparison between our results and the ones obtained 
by Pastore et al.~within a different framework.  
\end{itemize}
The final results of our work are summarized in Eqs.~(\ref{1pi_current_final}), 
(\ref{1pi_charge_final}), (\ref{con2})  and (\ref{con3}) which contain the expressions 
for the one-pion exchange
and short-range contributions to the two-nucleon current and charge densities at order $eQ$ (leading
loop order). It would be interesting to explore effects of these novel contributions to the 
exchange current and charge densities in e.g.~electron scattering off light nuclei. This work is in progress, 
see also \cite{Rozpedzik:2011cx}
for a pioneering calculation along this line concentrating on the two-pion exchange contributions.

\section*{Acknowledgments}

We would like to thank  Daniel Phillips for useful comments on the
manuscript and Walter Gl\"ockle, Jacek Golak, Dagmara Rozpedzik and
Henryk Wita{\l}a for many stimulating discussions on this topic.  S.K.~would
also like to thank Jacek Golak and Henryk Wita{\l}a for their hospitality
during his stay in Krakow where a part of this work was done. 
This work was
supported by funds provided by the Helmholtz Association (grants VH-NG-222 and
VH-VI-231), by the DFG (SFB/TR 16 ``Subnuclear
Structure of Matter''), by the EU HadronPhysics2 project ``Study
of strongly interacting matter'' and the European Research Council
(ERC-2010-StG 259218 NuclearEFT). 

\appendix

\def\theequation{\Alph{section}.\arabic{equation}}
\setcounter{equation}{0}
\section{Hamilton density}
\label{append_Ham}

In this appendix we define the expressions for the Hamilton density and currents. First let
us recapitulate the expressions we already defined in~\cite{Kolling:2009iq}
\begin{eqnarray}
\label{vertices}
\mathcal{H}_{21}^{(1)} &=&  \frac{\mathring{g}_A}{2F} N^\dagger \left(\vec{\sigma}
      \vec{\tau}\cdot\cdot\vec{\nabla}\vec{\pi} \right) N \,, \nn
\mathcal{H}_{22}^{(2)} &=&  \frac{1}{4 F^2}  N^\dagger
  \big[\vec{\pi}\times\dot{\vec{\pi}} \big]\cdot \vec{\tau} N\,, \nn
\mathcal{H}_{42}^{(4)} &=&  \frac{1}{32 F^4} \left(N^\dagger
    \left[\vec{\tau}\times\vec{\pi} \right] N \right) \cdot \left(N^\dagger
    \left[\vec{\tau}\times\vec{\pi} \right] N \right)
\,, \nn 
{J^0_{20}}^{(-1)} &=&\frac{e}{2} N^\dagger \left(\one
      +\tau_3 \right)N  = e N^\dagger \, \hat{e}\, N\,, \nn
{J_{02}^0}^{(-1)} &=& e  \big[\vec{\pi}
    \times  \dot{\vec{\pi}} \big]_3 \,, \nn
{\vec J_{02}\,}^{(-1)} &=& - e \big[\vec{\pi}
    \times \vec{\nabla} \vec{\pi} \big]_3 \,, \nn
{\vec J_{21}\,}^{(0)} &=& e\,\frac{
    \mathring{g}_A }{2F}  N^\dagger         \vec{\sigma} \left[ \vec{\tau} \times \vec{\pi}\right]_3
    N\,.
\end{eqnarray}

The definitions of the other parts of the Hamiltonian density and currents is given below.
\begin{eqnarray}
\mathcal{H}_{23}^{(3)} & = & \frac{\mathring{g}_A}{4F^3}
N^\dagger\left[2\, \vec{\sigma}\vec{\tau}\cdot\cdot\vec{\nabla} \vec{\pi} \,\vec{\pi}^{\,
    2}\,  \xi  + (1-4\xi) 
  \vec{\tau}\cdot\vec{\pi} \, \vec{\sigma}\cdot\vec{\nabla} \vec{\pi}^{\, 2}
 \right] N\,,\nn
\mathcal{H}_{04}^{(2)} & = & \frac{1}{8F^2}
N^\dagger\left[ 8\xi \, \partial_\mu \vec{\pi} \cdot\partial^\mu \vec{\pi} \,
  \vec{\pi}^{\, 2} - (1-4\xi) \left(\vec{\pi}
    \cdot \partial_\mu\vec{\pi}\right)^2 - (8\xi-1) M_\pi^2 \,
  \vec{\pi}^{\, 4}\right]\,,\nn
\mathcal{H}_{04}^{(2)} & = & \frac{C_S}{2} N^\dagger  N \, N^\dagger  N +
\frac{C_T}{2} N^\dagger \, \vec{\sigma} N \cdot N^\dagger \,\vec{\sigma}
N\,,\nn
\mathcal{H}_{21}^{(3)} & = & \frac{2d_{16} - d_{18}}{F}M_\pi^2\,
N^\dagger\vec{\sigma}\vec{\tau}\cdot\cdot\vec{\nabla} \vec{\pi}N\,,\nn
\tilde{\mathcal{H}}_{21}^{(3)} & = & -\frac{\mathring{g}_A i}{4 m_N  F} N^\dagger
\vec{\sigma}\cdot\overleftrightarrow{\nabla} \, \vec{\tau} \cdot \dot{\vec{\pi}} N\,,\nn
\tilde{\mathcal{H}}_{20}^{(2)} & = & -\frac{1}{2m_N}N^\dagger \vec{\nabla}^{\,
2} N\,, \nn
{J^0_{02}}^{(1)} & = & - 2 i e \frac{l_6}{F^2} \epsilon_{3 a b} \,
\vec{k}\cdot\vec{\nabla}\pi_a\, \dot{\pi}_b\,,\nn
{\vec{J}_{02}\, }^{(1)} & = &  2 i e  \epsilon_{3 a b} \,\left(\frac{l_6}{F^2} \, \vec{\nabla}
\pi_a \, \vec{k}\cdot\vec{\nabla}\pi_b  +i l_4
\frac{M_\pi^2}{F^2}\pi_a \vec{\nabla}\pi_b\right)\,,\nn
{J^0_{20}\, }^{(1)} & = &  e \vec{k}^{\, 2} N^\dagger (d_6 \tau_3 + 2 d_7)N\,,\nn
{\tilde{\vec{J}}_{20}}^{(1)} & = & -\frac{i e}{4m_N} N^\dagger\left[ (\one  + \tau^3)
  \overleftrightarrow{\nabla} + (1+c_6) \,(\one  +
  \tau^3) \vec{\sigma} \times \vec{k}  + 2 c_7\,
  \vec{\sigma}\times\vec{k}\right] N\,,\nn
{\tilde{\vec{J}}_{20}}^{(1)} & = & -\frac{i e}{4m_N} N^\dagger
(\one  + \tau^3)
  \overleftrightarrow{\nabla} 
     N\,,\nn
{J^0_{21}}^{(2)} & = &  - e\frac{i}{F} N^\dagger \epsilon_{3 a b} \tau_b  (d_{20} + d_{21} - \frac{d_{22}}{2}) \vec{\sigma}\cdot\vec{k}
\, \dot{\pi}_a   N\,,\nn
{\vec{J}_{21}\,}^{(2)} & = & \frac{e}{F} N^\dagger\biggl[
- 4 i d_{8} \vec{k} \times \vec{\nabla}\pi_3- 4 i d_{9}
\vec{k} \times \vec{\nabla}\pi_a \, \tau_a +  \epsilon_{3 a b}
\tau_b  \biggl( -\vec{\sigma} \pi_a (2d_{16} -
d_{18}) \,M_\pi^2  \nn 
&&{}
 + i \left(d_{21} - \frac{d_{22}}{2}\right)
 \,\vec{k} \times\left[\vec{\sigma}\times \vec{\nabla}\right] \pi_a - \frac{d_{22}}{2}
 \pi_a   \vec{k}\times\left[
\vec{k}\times\vec{\sigma}\right] 
 \biggr)\biggr]N\,,\nn
{\tilde{\vec{J}}_{21}}^{(2)} & = & -\frac{e \mathring{g}_A}{4m_NF} N^\dagger
\vec{\sigma} (\tau_a  + \delta_{a3}) \dot{\pi}_a
 N\,.
\end{eqnarray}

\def\theequation{\Alph{section}.\arabic{equation}}
\setcounter{equation}{0}
\section{Loop integrals}
\label{app2}

The following integrals contribute to the one-pion exchange current 
operator at the leading one-loop order:
\begin{eqnarray}
\int \frac{d^3l}{(2\pi)^3} \frac{1}{\omega_l} & = & 2\Delta_\pi\,, \nn
\int \frac{d^3l}{(2\pi)^3} \frac{l^2}{\omega_l^3} &=& 6 \Delta_\pi\,, \nn
\int \frac{d^3l}{(2\pi)^3} \frac{l^2}{\omega_l^4} &=& - \frac{3M_\pi}{8\pi}\,,
\nn
\int \frac{d^3l}{(2\pi)^3}\frac{l^a\, l^b}{\omega_+
  \omega_-(\omega_++\omega_-)} & \equiv & \mathcal{I}_2 \, \delta^{ab}+
k^ak^b\, \mathcal{I}_3\,,\nn
\int \frac{d^3l}{(2\pi)^3}
\frac{\omega_--\omega_+}{\omega_+\omega_-(\omega_++\omega_-)^2}l^a &
\equiv & k^a\, \mathcal{K}_1\,,\nn
\int \frac{d^3l}{(2\pi)^3}
\frac{2\omega_++\omega_-}{\omega_-\omega_+^3(\omega_++\omega_-)^2}l^a
& \equiv & k^a \mathcal{B}_1\,, \nn
\int \frac{d^3l}{(2\pi)^3}
\frac{2\omega_++\omega_-}{\omega_-\omega_+^3(\omega_++\omega_-)^2}l^al^b
 &\equiv&  \delta^{ab} \mathcal{B}_2 +k^ak^b \mathcal{B}_3\,,\nn
\int \frac{d^3l}{(2\pi)^3}
\frac{2\omega_++\omega_-}{\omega_-\omega_+^3(\omega_++\omega_-)^2}l^al^bl^c
& \equiv & (\delta^{ab}k^c+\delta^{ac}k^b+\delta^{bc}k^a) \mathcal{B}_4+
k^ak^bk^c\mathcal{B}_5\,,\nn
\int \frac{d^3l}{(2\pi)^3}\frac{1}{\omega_+^2\omega_-^2} & \equiv &
\mathcal{A}_0\,, \nn
\int \frac{d^3l}{(2\pi)^3}\frac{l^a\, l^b}{\omega_+^2\omega_-^2} &\equiv& 
\mathcal{A}_2\delta^{ab}+ \mathcal{A}_3 k^a\, k^b \,,
\end{eqnarray}
where the pion energies $\omega_l$ and $\omega_\pm$ are defined as
\beq
\omega_l = \sqrt{l^2+M_\pi^2}\,, \qquad 
\omega_\pm = \sqrt{\left(\vec{l}\pm\vec{k}
  \right)^2+4M_\pi^2}\,.
\eeq
The integrals can be computed explicitly in dimensional regularization. We
only give here the results for the integrals that are actually needed. These are:
\begin{eqnarray}
 \mathcal{I}_2 & = & \left(4 + \frac{2k^2}{3M_\pi^2}\right) \Delta_\pi+
 \frac{s^2L(k)}{12 \pi^2}  - \frac{5k^2 +   24M_\pi^2}{72\pi^2} \,,\nn
\mathcal{I}_3 & = & - \mathcal{K}_1  = - 2 \mathcal{B}_4 =  -\frac{2
  \Delta_\pi}{3M_\pi^2} - \frac{s^2 L(k)}{12\pi^2k^2} +
\frac{5k^2+24M_\pi^2}{72\pi^2k^2}\,, \nn
\mathcal{B}_1 & = & -\mathcal{B}_3 = - \frac{L(k)-1}{8\pi^2 k^2} \,,\nn
\mathcal{B}_2 & = & - \frac{\Delta_\pi}{M_\pi^2}  - \frac{2L(k)-1}{16\pi^2} \,,\nn
\mathcal{B}_5 & = & -\frac{(4M_\pi^2 + k^2)L(k)}{8\pi^2k^4}  + \frac{3M_\pi^2+k^2}{6\pi^2k^4} \,,\nn
\mathcal{A}_0 & = & \frac{A(k)}{4\pi}\,,\nn
\mathcal{A}_2  &=& -\frac{M_\pi+s^2A(k)}{8\pi} \,,\nn
\mathcal{A}_3 &=& -\frac{M_\pi - s^2 A(k)}{8\pi k^2}\,.
\end{eqnarray}
where we have introduced 
\beq
\label{loop_functions}
  L(k)  =  \frac{1}{2}\frac{s}{k}\log\left(\frac{s+k}{s-k} \right), \qquad
  A(k) = \frac{1}{2k}\arctan\left(\frac{k}{2M_\pi}\right),\qquad
    s  =  \sqrt{k^2 + 4M_\pi^2},
\eeq
with $k = |\vec{k}\,|$. Further, the integral $\Delta_\pi$ is defined in Ref.~\cite{Bernard:1995dp}
according to 
\beq
\label{def_delta_pi}
\Delta_\pi =  \lim_{d \to 4}\frac{1}{i}\int \frac{d^d l}{\left(
    2\pi\right)^d}\frac{1}{M_\pi^2 - l^2- i \epsilon}
 =  2M_\pi^2\left(\lim_{d \to 4}L + \frac{1}{16\pi^2}\log\left(\frac{M_\pi}{\mu} \right)
 \right) 
, 
\eeq
where the quantity $L$ has a pole in $d=4$ dimensions and is given by  
\beq
\label{def_L}
  L  =\frac{\mu^{d-4}}{16 \pi^2}\left[ \frac{1}{d-4} -
    \frac{1}{2}\left(\Gamma^\prime(1) + 1 + \log\left(4\pi \right) \right)\right].
\eeq

\def\theequation{\Alph{section}.\arabic{equation}}
\setcounter{equation}{0}
\section{Configuration-Space Expressions}
\label{app3}

For the sake of completeness, we also give the expressions in configuration space 
obtained by carrying out the Fourier-transformation of the momentum space results
\begin{equation}
\mathcal{F} \left( f (\vec q_1 \,, \, \vec q_2 \,) \right) \equiv
\int \frac{d^3q_1}{(2\pi)^3} \frac{d^3q_2}{(2\pi)^3} e^{i
  \vec{q}_1 \cdot \vec{r}_{1}}  e^{i  \vec{q}_2 \cdot \vec{r}_{2}}\, f (\vec
q_1 \,, \, \vec q_2 \,) \,(2\pi)^3\, \delta^3\left( \vec{q}_1 + \vec{q}_2 - \vec{k}
\right)\, .
\end{equation}
We find the following expressions:
\begin{eqnarray}
\label{1pi_current_final_conf}
  \vec{J}_{1\pi}  & = &\biggl(- \left[ 
   \tau_2^3 \, f_1( k ) +   \vec{\tau}_1 \cdot\vec{\tau}_2 \, f_2( k
   ) \right]\,
 \vec{\sigma}_2\cdot\vec{\nabla}_{12} \, \left[ \vec{k}
   \times\vec{\nabla}_{12}\right]  + 
 \left[\vec{\tau}_1\times\vec{\tau}_2 \right]^3
  \vec{\sigma}_2\cdot\vec{\nabla}_{12}  \biggl[  - 
  \vec{k}\times\left[\vec{\nabla}_{12}\times\vec{\sigma}_1 
    \right] f_3( k)\, \nn
 &+&
   \vec{k}\times\left[\left(\vec{\nabla}_{12} \,  +i
       \vec{k}\right)\times\vec{\sigma}_1\right] \, f_4(k ) +  \vec{\sigma}_1 \cdot
    \left(\vec{\nabla}_{12} + i \vec{k}\right)\,  \frac{\vec{k}}{k^2}\, f_5(
      k ) - i \vec{\sigma}_1\,  f_6( k )
  \biggl]\biggl) \, \frac{M_\pi e^{-M_\pi\, r_{12}}}{4\pi \,
     r_{12}}e^{i \vec{k}\cdot\vec{r}_1}
  \nn
&+& \frac{i}{2}\, \left[\vec{\tau}_1 \times \vec{\tau}_2 \right]^3
  \vec{\sigma}_2\cdot \left(\vec{\nabla}_{12} - \frac{i\vec{k}}{2} \right)\,  \vec{\sigma}_1\cdot
      \left(\vec{\nabla}_{12} + \frac{i\vec{k}}{2} \right) \,\biggl[ 
   f_5( k )- f_6(k )
\biggl]\, \vec{\nabla}_{12} \, f\left(\vec{k},\vec{r}_{12} \right)  e^{i \vec{k}\cdot\vec{R}} \, ,
  \end{eqnarray}
where we have introduced
\begin{eqnarray}
  \vec{r}_{12} & = & \vec{r}_1 - \vec{r}_2\, , \qquad \vec{R} = \frac{\vec{r}_1 +
  \vec{r}_2}{2}\, , \qquad \vec{\nabla}_{12} = \vec{\nabla}_{r_{12}}\, , \nn
f\left(\vec{k},\vec{r}_{12} \right) & = & \int \frac{d^3p}{(2\pi)^3} \int_{-1}^1 dx \, 
 \frac{e^{i \vec{r}_{12}\cdot\left( \vec{p} + \vec{k}\, x/2 \right)}}{(p^2+M_\pi^2
  + (1-x^2)k^2/4)^2}\,.
\end{eqnarray}
The one-pion exchange charge density has the following form:
\begin{eqnarray}
\label{1pi_charge_final_conf}
  \rho_{1\pi} & = &\biggl(-2 M_\pi^2\, 
  \vec{\sigma}_2\cdot\vec{\nabla}_{12}\tau_2^3\biggr[ 
  \vec{\sigma}_1\cdot\vec{k} \, \vec{\nabla}_{12} \cdot\vec{k}\, f_7( k
  ) + \vec{\nabla}_{12}\cdot\vec{\sigma}_1 \, f_8 (k )\biggl]
  \, + \, \frac{ie g_A^2\, M_\pi^2}{8F_\pi^2m_N}  \biggl[(1-2\bar\beta_9) \left(\tau_2^3 +
   \vec{\tau}_1\cdot\vec{\tau}_2 \right) \vec{\sigma}_1\cdot\vec{k}
 \vec{\sigma}_2\cdot\vec{\nabla}_{12} \nn
  & - & 
  i(1+2\bar\beta_9) \left[
   \vec{\tau}_1\times\vec{\tau}_2\right]^3
    \left(\vec{\sigma}_1\cdot\vec{k}_1
 \vec{\sigma}_2\cdot\vec{\nabla}_{12} - \vec{\sigma}_2\cdot\vec{k}_2
 \vec{\sigma}_1\cdot\vec{\nabla}_{12} \right)\biggr] - \frac{ie g_A^2
        }{16 m_N F_\pi^2}  \vec{\sigma}_1\cdot\vec{\nabla}_{12}  \,
        \vec{\sigma}_2\cdot\vec{\nabla}_{12}
      \, \biggl[(2\bar\beta_8-1)(\tau_2^3 + \left(
  \vec{\tau_1}\cdot\vec{\tau}_2\right)) \,   \nn
 &\times & \vec{\nabla}_{12}\cdot\vec{k} +  i \left[
  \vec{\tau}_1 \times \vec{\tau}_2\right]^3  
\biggl(-\left(
   2\bar\beta_8 + 1\right)
 \vec{\nabla}_{12}\cdot\vec{k}_2
+ \left( 2\bar\beta_8  -1 \, \right)    \,
  \vec{\nabla}_{12} \cdot\vec{k}_1
\,\biggr)\biggr]\biggr) \, \frac{e^{-M_\pi
    r_{12}}}{8 \pi\,M_\pi\,  r_{12}}e^{i \vec{k}\cdot\vec{r}_1}\nn
&+&e \frac{g_A^2  }{16 F_\pi^2 m_N}\,\left[\vec{\tau}_1\times\vec{\tau}_2\right]^3\, \vec{\sigma}_1\cdot\left(\vec{\nabla}_{12} +
  \frac{i \vec{k}}{2}\right)\, \vec{\sigma}_2\cdot\left(\vec{\nabla}_{12} -
  \frac{i \vec{k}}{2}\right)\, \left(\vec{\nabla}_{12} +
  \frac{i \vec{k}}{2} \right) \cdot\vec{k}_1\,f\left(\vec{k},\vec{r}_{12}
\right) \, e^{i \vec{k}\cdot\vec{R}}\,. 
\end{eqnarray}

Finally, we also give the coordinate-space expression for the
short-range currents:
\begin{eqnarray}
\rho_{\rm contact} &=&   C_T\, \tau_1^3\, \left[
      \vec{\sigma}_1\cdot\vec{k}\,\vec{\sigma}_2\cdot\vec{k}\,\frac{2
      g_A^2}{F_\pi^2}\, f_{9}\left(k \right) + \vec{\sigma}_1\cdot\vec{\sigma}_2 \,
    f_{10}\left( k\right)\right] e^{i \vec{k} \cdot\vec{R}}
\delta^3\left( \vec{r}_{12}\right)\,,\\
 \vec{J}_{\rm contact} & = & \biggl\{e\,\frac{1}{16}\left[ \vec{\tau}_1 \times \vec{\tau}_2\right]^3 \,\biggl[\left(C_2 +3C_4 + C_7\right)
    \vec{\nabla}_{12} - \left(-C_2 + C_4 + C_7 \right) \,
  \, \left(\vec{\sigma}_1 \cdot\vec{\sigma}_2 \right)\, \vec{\nabla}_{12} \nn
   &+& C_7 \, \left(  \vec{\sigma}_2\cdot
     \vec{\nabla}_{12} \,  \vec{\sigma}_1+ \, \vec{\sigma}_1\cdot
     \vec{\nabla}_{12}\,  \vec{\sigma}_2 \right)\biggr] 
- e\,\frac{C_5 \, }{16}\,  \tau_1^3\,
 \left[\left(\vec{\sigma}_1 + \vec{\sigma}_2 \right)\times
     \vec{\nabla}_{12} \right] 
+ i e L_1 \,\tau_1^3 \,
\left[\left(\vec{\sigma}_1 - \vec{\sigma}_2 \right)\times\vec{k}\right] \nn
&+& i e L_2 \, \left[\vec{\sigma}_1 \times\vec{k} \right]\biggr\}  e^{i \vec{k} \cdot\vec{R}}
\delta^3\left( \vec{r}_{12}\right)\, . 
\nonumber
\end{eqnarray}



\end{document}